\documentclass[fleqn,usenatbib]{mnras}

\usepackage[T1]{fontenc}

\usepackage{graphicx}	% Including figure files
\usepackage{amsmath}	% Advanced maths commands
\usepackage{amssymb}	% Extra maths symbols
\usepackage{url,times,graphicx,hyperref,amsmath,amsfonts,amssymb,aas_macros,color,epsfig,epstopdf}
\usepackage{booktabs}
\usepackage[utf8]{inputenc}
\usepackage{ae,aecompl}

\newcommand{\kms}            {\,{\rm km\,s^{-1}}}
\newcommand{\vmax}           {V_{\rm max}}

\newcommand{\msun}{M$_\odot$}
\title[Satellites and the faint galaxy mass-halo mass relation]{
Satellite mass functions and the faint end of the galaxy mass-halo mass relation in LCDM
}

\author[I. Santos-Santos et al.]{
Isabel M.E. Santos-Santos,$^{1}$\thanks{E-mail: isantos@uvic.ca}
A. N. Other,$^{2}$
Third Author$^{2,3}$
and Fourth Author$^{3}$
\\
$^{1}$Department of Physics and Astronomy, University of Victoria, Victoria, BC V8P 5C2, Canada\\
}

\author[Santos-Santos et al.]{
\parbox[t]{\textwidth}{
Isabel M.E. Santos-Santos$^{1,2}$\thanks{E-mail: isabel.santos@durham.ac.uk}, 
Laura V. Sales$^{3}$, 
Azadeh Fattahi$^{2}$ and 
Julio F. Navarro$^{1}$
}\\
\\
$^{1}$Department of Physics and Astronomy, University of Victoria, Victoria, BC V8P 5C2, Canada\\
$^{2}$Institute for Computational Cosmology, Department of Physics, Durham University, South Road, Durham, DH1 3LE, UK\\
$^{3}$Department of Physics and Astronomy, University of California Riverside, 900 University Avenue, CA 92507, USA\\
}

\pubyear{2020}

\date{Accepted XXX. Received YYY; in original form ZZZ}

\pubyear{2020}

\begin{document}
\label{firstpage}
\pagerange{\pageref{firstpage}--\pageref{lastpage}}
\maketitle

\begin{abstract}
  The abundance of the faintest galaxies provides insight into the nature of dark matter and the process of dwarf galaxy formation. In the LCDM scenario, low mass halos are so numerous that the efficiency of dwarf formation must decline sharply with decreasing halo mass in order to accommodate the relative scarcity of observed dwarfs and satellites in the Local Group. The nature of this decline contains important clues to the mechanisms regulating the onset of galaxy formation in the faintest systems. We explore here two possible models for the stellar mass ($M_*$)-halo mass ($M_{200}$) relation at the faint end, motivated by some of the latest LCDM cosmological hydrodynamical simulations. One model includes a sharp mass threshold below which no luminous galaxies form, as expected if galaxy formation proceeds only in systems above the Hydrogen-cooling limit. In the second model, $M_*$ scales as a steep power-law of $M_{200}$ with no explicit cutoff, as suggested by recent semianalytic work. Although both models predict satellite numbers around Milky Way-like galaxies consistent with current observations, they predict vastly different numbers of ultra-faint dwarfs and of satellites around isolated dwarf galaxies. Our results illustrate how the satellite mass function around dwarfs may be used to probe the $M_*$-$M_{200}$ relation at the faint end and to elucidate the mechanisms that determine which low-mass halos ``light up'' or remain dark in the LCDM scenario.
\end{abstract}


%%%%%%%%%%%%%%%%%%%%%%%%%%%%%%%%%%%%%%%%%%%%%%%%%%

%%%%%%%%%%%%%%%%% BODY OF PAPER %%%%%%%%%%%%%%%%%%

\section{Introduction}\label{sec:intro}

Ultrafaint dwarfs, defined here as dwarf galaxies with stellar masses $M_* < 10^5$ \msun \citep{Bullock2017}, are typically systems whose extremely low surface brightness ($\mu_v \geq 27$ $\rm mag/\rm arcsec^2$) hinders their discovery and makes follow-up studies extremely difficult. Indeed, although recent efforts have led to the discovery of dozens of ultrafaints in the Milky Way (MW) halo \citep[see ][ and references therein]{Simon2019}, it remains unclear how many more of them may still lurk undetected in the vicinity of our Galaxy.

%Suspected to be highly dark matter dominated \citep{McConnachie2012}, their low stellar content also means that measurements of their internal kinematics are extremely challenging.
The ultrafaint population also remains largely unexplored in external galaxies, with only loose constraints available on the massive-end of the ultrafaint regime in M31
\citep[see the dwarf galaxy catalog  compiled and maintained by][]{McConnachie2012}.
 Identifying isolated ultrafaints in the field is even more difficult, with few, if any, reported so far outside the Local Group. 
 % none reported so far outside the Local Volume (??define Local Volume??). 

Because of their extreme intrinsic faintness, few bright stars are available for spectroscopic study in ultrafaints, even when using some of the largest ground-based telescopes. This implies that the characterization of some of their basic properties, such as their metallicity distribution, elemental abundances, or velocity dispersion, is subject to large uncertainty. Poorly determined velocity dispersions, in particular, affect our ability to estimate halo masses and to constrain the relation between stellar mass ($M_*$) and halo virial\footnote{We shall use halo ``virial'' properties defined within a radius, $r_{200}$, enclosing a mean density 200 times the critical density for closure. A subscript ‘200’ identifies quantities defined within or at that radius. } mass ($M_{200}$) at the very faint end of the galaxy luminosity function.

Indeed, our best constraints on the $M_*$-$M_{200}$ relation at the faint end arguably comes from abundance-matching techniques \citep{Conroy2006,Guo2010,Moster2013,Behroozi2013}. Because the galaxy stellar mass function around $M_{*} \sim 10^8\, M_\odot$ \citep[the faintest luminosities for which it is well constrained; see, e.g.,][]{Baldry2012} is substantially shallower than the LCDM halo mass function in that regime
\citep{Springel2008,BoylanKolchin2009},
 it is clear that galaxy formation must become increasingly inefficient towards decreasing halo masses. Characterizing this decline in galaxy formation efficiency at the low-mass end is difficult, and there is so far no consensus on how steep the decline is, on what the scatter in $M_*$ at fixed $M_{200}$ might be, and on whether there is a characteristic ``threshold'' halo mass below which no luminous galaxy forms in LCDM.

The lack of consensus concerns not only abundance-matching studies, but also direct cosmological simulations of the formation of the faintest galaxies.  For example, Local Group simulations from the APOSTLE project \citep{Sawala2016,Fattahi2016} suggest a relation with a fairly sharp cutoff at low halo masses, where few, if any, isolated halos with $\vmax$ below $\sim15 \kms$ host a galaxy \citep{Fattahi2018}. At least qualitatively, this is the behaviour expected in scenarios where luminous galaxy formation only proceeds in halos with masses exceeding the ``hydrogen-cooling limit'' (HCL) set by the primordial abundance cooling function after accounting for the presence of an evolving, ionizing UV background \citep[see; e.g.,][]{Gnedin2000,Okamoto2008,BenitezLlambay2020}.

On the other hand, some cosmological simulations suggest that even halos below the HCL may be able to form stars, so that no clear minimum ``threshold mass'' for galaxy formation exists. For example, FIRE-2 simulations \citep{Hopkins2018, Wetzel2016, Wheeler2019} 
%or those from the Marvel/DC Justice league collaboration \citep{Munshi2021} 
seem better described by a power-law $M_*$-$M_{200}$ relation similar to that reported by \citet{Brook2014} and which extends well below the HCL mass.

This argument has been strengthened by semi-analytic models that attempt to reproduce simultaneously the MW satellite mass function and its radial distribution. Because tides may, in principle, disrupt subhalos near the MW disk, accounting for the large number of ultrafaint satellites discovered in the inner $\sim 40$ kpc of the MW halo has led to the suggestion that populating subhalos well below the HCL with luminous galaxies may be needed \citep{Kelley2019, Graus2019}. % Nadler2020}.

However, there is still substantial uncertainty about whether Galactic tides are actually able to fully disrupt cuspy LCDM subhalos \citep{vandenBosch2018,Errani2021} and no cosmological simulation has actually reached the ultrafaint regime probed by observations. Despite these uncertainties, it is clear that simulation predictions for the faintest dwarfs appear to differ, depending on the resolution and subgrid physics adopted in the simulations \citep[see, e.g.,][]{Munshi2019}.  This is problematic, as the steep halo mass function in LCDM implies that even small differences in the stellar mass-halo mass relation should result in large differences in the expected number of faint galaxies.

We explore here how the abundance of ultrafaint satellites may be used to place constraints on the behaviour of the $M_*$-$M_{200}$ relation at the faint end. Their abundance around isolated dwarf primaries is particularly constraining.  This is because the subhalo mass function is well approximated by a power law \citep{Springel2008} and, therefore, a power-law stellar mass-halo mass relation would result in ``self-similar''  satellite mass functions independent of primary mass \citep{Sales2013}. This is a clear prediction that can be used to gain insight into the shape of the stellar mass-halo mass relation for primaries at the faint end.

We  explore these issues here, and argue that  the ultrafaint satellites of isolated dwarf galaxies is a promising way to elucidate how the faintest galaxies form and populate dark halos at the low mass end. This paper is organized as follows. We begin  by motivating in Sec.~\ref{sec:model} two particular analytic forms of the faint-end $M_*$-$M_{200}$ relation (a power-law and one with an explicit low-mass cutoff)  based on results from recent cosmological hydrodynamical simulations (Sec.~\ref{sec:methods}). We validate the ``cutoff'' model in Sec. ~\ref{ssec:compareAP} 
%using our model to reproduce
by reproducing
 results from the APOSTLE runs. We then compare the results from both our models for the ultrafaint satellite population of dwarf primaries spanning a wide range of stellar mass (Sec.~\ref{ssec:compare}), and then contrast these results with available data for the Local Group in Sec. ~\ref{ssec:data}. We conclude with predictions for future satellite surveys of ultrafaint dwarfs around primaries such as the Large Magellanic Cloud (LMC)
 %Large and Small Magellanic Clouds (LMC and SMC, respectively) primaries
  in Sec.~\ref{ssec:data_future} and summarize our main results in Sec. ~\ref{sec:conclu}.%??This paragraph needs rewriting??

\section{Numerical Methods}\label{sec:methods}

We shall use results from a number of recent cosmological hydrodynamical simulations of dwarf galaxy formation in LCDM. These include simulations of individual galaxies from the NIHAO project \citep{Wang2015,Buck2019}, an ensemble of simulations using the FIRE \citep{Hopkins2018,Wheeler2019,GarrisonKimmel2019} and CHANGA \citep{Munshi2021} codes, as well as simulations of constrained Local Group environments from the APOSTLE project \citep{Sawala2016,Fattahi2016,Fattahi2018}. Since we shall use the latter to calibrate our modeling procedure we describe the APOSTLE simulations in some detail below. Results from the other runs are taken directly as reported in those publications, to which we refer the interested reader for details.

\subsection{The APOSTLE simulations}\label{sec:sims}

%We use the APOSTLE cosmological hydrodynamical simulations to calibrate our models.
The APOSTLE project is a set of 12 `zoom-in' cosmological volumes tailored to reproduce the main properties of the Local Group. Each volume is selected from a large cosmological box to contain a pair of halos with masses, relative radial and tangential velocities, and surrounding Hubble flow, consistent with the corresponding values observed for the Milky Way-Andromeda pair \citep{Fattahi2016}.

The APOSTLE runs used the EAGLE galaxy formation code \citep{Schaye2015,Crain2015}, using the so-called `Reference' parameters. This code includes subgrid physics recipes for radiative cooling, star formation  in gass exceeding a metallicity-dependent density threshold, stellar feedback from stellar winds,  radiation pressure and supernovae explosions, homogeneous X-ray/UV background radiation, supermassive black-hole growth and AGN feedback (note that the latter has negligible effects on dwarf galaxies and is therefore unimportant in APOSTLE).

The EAGLE model was calibrated to approximate the observed $z=0.1$ galaxy stellar mass function in the $M_*=10^8$-$10^{12}$ M$_\odot$ range.  Simulated galaxies thus roughly follow the abundance-matching $M_*$-$M_{200}$ relation of \citet{Behroozi2013} and \citet{Moster2013}. No extra calibration is made in APOSTLE, and therefore the stellar-halo mass relation that results for fainter galaxies may be regarded as the extrapolation of the same subgrid physics to lower mass halos.

The APOSTLE volumes have been run at three different levels of resolution. In this paper we use the $5$  highest-resolution volumes \citep[labelled "AP-L1" in][]{Fattahi2016}. These runs have initial dark matter and gas particle masses of $m_{\rm DM}\sim 5\times 10^4$ M$_\odot$  and $m_{\rm gas}\sim 1\times 10^4$ M$_\odot$, respectively, and a gravitational softening length of $134$ pc at $z=0$. The APOSTLE volume simulated at highest resolution fully contains a sphere of  radius $r\sim 3.5$ Mpc from the midpoint of the MW and M31 analog halos.
 
The friends-of-friends (FoF) groupfinding algorithm \citep{Davis1985} (with linking length equal to 0.2 times the mean interparticle separation) and the SUBFIND halo finder \citep{Springel2001, Dolag2009} were used to identify haloes and subhaloes.  We shall refer to the galaxies formed in the most massive subhalos of each FoF group as ``centrals'', and to the rest of galaxies within the virial radius of each FoF central as ``satellites''. Throughout the paper we shall use the term ``primary'' to refer to a central galaxy that may have satellites.

APOSTLE assumes a flat LCDM cosmological model following WMAP-7 parameters \citep{Komatsu2011}: $\Omega_{\rm m}=0.272$; $\Omega_\Lambda=0.728$; $\Omega_{\rm bar}=0.0455$; $H_0=100 \, h$ km s$^{-1}$ Mpc$^{-1}$; $\sigma=0.81$; $h=0.704$.

%a prediction from the simulation, and describes  a sharp 'cutoff' at low halo masses 
%or $V_{\rm max}$ 
%such that in general all isolated dwarf galaxies inhabit halos with virial mass  
%$V_{\rm max}>20$ km/s   
%$M_{\rm 200}\gtrsim 10^{9}$ M$_\odot$
%  \citep{Fattahi2016b,Fattahi2018}. Here, we use virial quantities defined at $r_{200}$, the radius enclosing an average density equal to $200$ the critcal density of the universe. 

\section{Modeling the satellite stellar mass function}\label{sec:model}

The satellite mass function of a primary of given stellar mass, $M_*^{\rm pri}$, depends mainly on (i) the mass function of subhalos present in the halo of that system, on (ii) the relation between stellar mass and subhalo mass, and on (iii) the possible reduction of stellar mass due to tidal stripping after infall. The first item depends mainly on the primary halo virial mass, or, equivalently, on $V_{200}^{\rm pri}$, and has been extensively studied through cosmological N-body simulations.

For the second item, which, in the case of satellites, applies before first infall into the primary halo, it is customary to express the stellar mass not as a function of (sub)halo mass, but rather in terms of its maximum circular velocity, $V_{\rm max}$, a quantity more resilient to tidal effects than virial mass.

The third item is the most difficult to treat analytically, since it depends strongly on the pericentric distance of the orbit, the number of orbits completed, and the radial segregation of stars within each subhalo. Fortunately, as we shall see below, the fraction of stellar mass lost to tides is, on average, small, and we shall neglect it in our modeling in the interest of keeping the model as simple as possible.

We describe below the parametrizations we adopt to build an analytical model for the satellite mass function of a primary of mass $M_*^{\rm pri}$. These parametrizations are motivated by the results of cosmological N-body and hydrodynamical simulations, as discussed in detail in the remainder of this section. We note that the two satellite mass function models explored here differ only in the assumptions made for the $M_*$-$V_{\rm max}$ relation.

%%%%%%%%%%%%%%%%%%%%%%%%%%%%%%%%%%
\begin{figure}
\includegraphics[width=1\linewidth]{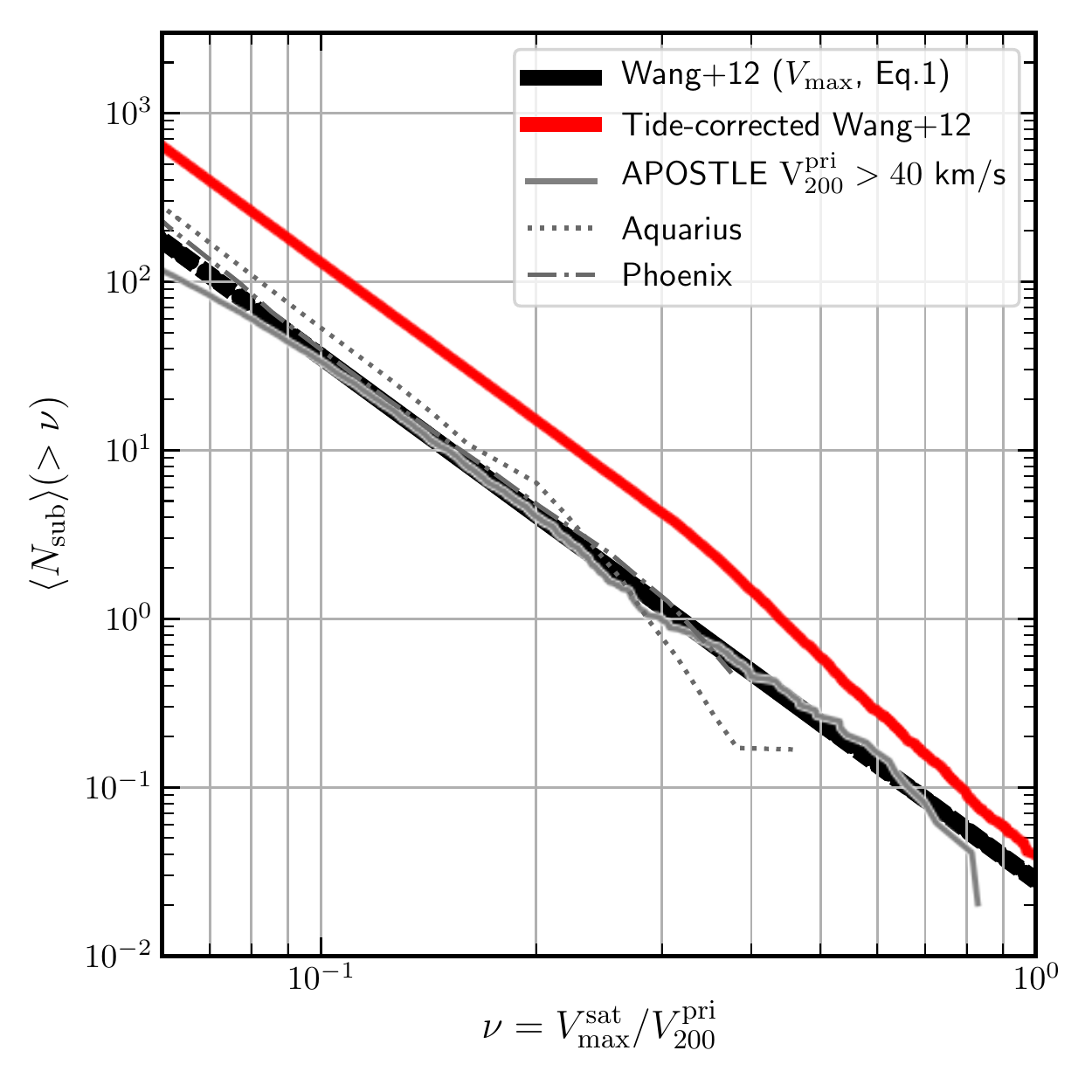}
\caption{The subhalo velocity function (i.e., average number of subhalos with $\nu=V_{\rm max}/V_{200}$ above  a certain value). The solid black line shows the function proposed by \citet{Wang2012} (Eq.~\ref{eq:wang}), which describes  well the substructure mass function of LCDM halos of all masses in the Millenium DM-only cosmological simulation, particularly in the $0.1<\nu<0.5$ range.  Extrapolations of this line beyond such range are shown in dotted line style.  The average subhalo velocity function for APOSTLE halos with $V_{\rm 200}^{\rm pri}>40$ km/s is shown with a solid gray line.  For comparison, thin gray dotted and dashed-dotted lines show the average subhalo functions found for halos in simulations of the Aquarius and Phoenix projects, respectively.  A thick red line shows the 'tide-corrected' version of Eq.~\ref{eq:wang}, resulting from statistically converting $V_{\rm max}$ values to $V_{\rm peak}$ as explained in Sec.~\ref{sec:model}.  
%%%The actual $V_{\rm peak}/V_{200}$ distribution for APOSTLE subhalos that are satellites of the MW or M31 analog primaries is shown with a dashed black line.
}
 \label{fig:subhaloVF}
\end{figure}
%%%%%%%%%%%%%%%%%%%%%%%%%%%%%%%%%%

%%%%%%%%%%%%%%%%%%%%%%%%%%%%%%%%%%
\begin{figure*}
\includegraphics[width=0.9\linewidth]{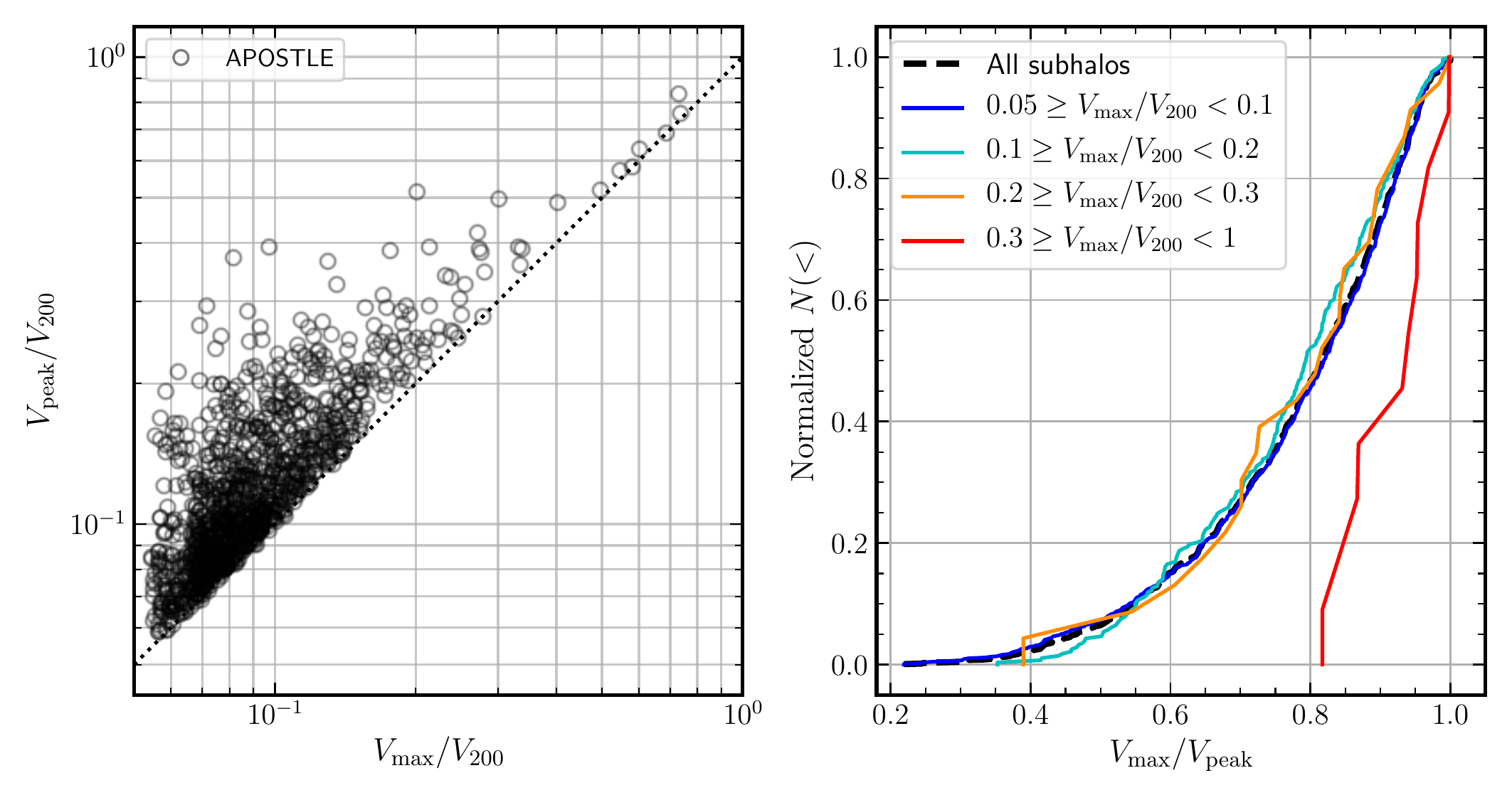}
\caption{
  $V_{\rm max}/V_{200}$  vs. $V_{\rm peak}/V_{200}$ (left) and normalized cumulative distribution of $V_{\rm max}/V_{\rm peak}$ (right) for APOSTLE subhalos. Lines of different colors show results for subhalos in different $V_{\rm max}/V_{200}$ bins, as indicated in the legend.
}
 \label{fig:vmaxvpeak}
\end{figure*}

%%%%%%%%%%%%%%%%%%%%%%%%%%%%%%%%%%

\subsection {Subhalo mass function}

%We assume that the substructure mass function of a given primary halo follows that predicted from $\Lambda$CDM  dark matter-only cosmological simulations which we have validated to reproduce also the subhalo velocity function of the hydrodynamical run in APOSTLE. For this we model the scaled subhalo velocity function of dark matter halos, i.e., the average number of dark matter subhalos expressed as a function of $\nu$, the maximum velocity ratio of the subhalo to that of the host.

The substructure mass function of LCDM halos scales in direct proportion to the virial mass of the primary halo and has been shown to be fairly well approximated by a power law. Following \citet{Wang2012}, the average number of subhalos  within the virial radius of an isolated (central) LCDM halo may be expressed as 
\begin{equation}\label{eq:wang}
\langle N_{\rm sub}\rangle(>\nu)=10.2\, (\nu/0.15)^{-3.11}
\end{equation}
where $\nu=V_{\rm max}/V_{200}^{\rm pri}$. This function applies to all LCDM halos regardless of mass, and has been tested well over the $0.1<\nu<0.5$ range.  The scatter around the average number at given $\nu$ is well approximated by Poisson statistics.

We compare in  Fig.~\ref{fig:subhaloVF} the results of three sets of cosmological simulations  with the predictions from Eq.~\ref{eq:wang} (thick black line). The simulations include the average of all Milky Way-sized halos of the Aquarius project \citep[dotted black line;][]{Springel2008}, that of the cluster-sized halos of the Phoenix project \citep[dot-dashed black line;][]{Gao2012}, as well as that of all halos with $V_{200}>40$ km/s in the APOSTLE project (solid grey line).
As is clear from this figure, Eq.~\ref{eq:wang} reproduces quite well the subhalo mass function of halos spanning a wide range of virial mass.

However, this function is expressed in terms of the present-day subhalo maximum circular velocity, $V_{\rm max}$, which may have been affected by tidal stripping after infall. Since the stellar content of a subhalo is more closely tied to $V_{\rm peak}$, the maximum circular velocity prior to infall, Eq.~\ref{eq:wang} must therefore be corrected to yield the distribution of $V_{\rm peak}$ values needed in the modeling.

To this end, we explore the relation between $V_{\rm max}$ and $V_{\rm peak}$ in APOSTLE halos. This  is shown in the left-hand panel of Fig.~\ref{fig:vmaxvpeak}, scaled by the virial velocity of the primary, $V_{200}$, at $z=0$. As expected,  APOSTLE subhalos had $V_{\rm peak}$ values systematically larger than $V_{\rm max}$. The distribution of the ratio $V_{\rm peak}/V_{\rm max}$ is shown (in cumulative form) in the right-hand panel of Fig.~\ref{fig:vmaxvpeak}, for various bins in $V_{\rm max}/V_{200}$.
This panel shows that, on average, the reduction in subhalo maximum circular velocity  that result from tides is fairly modest and largely independent of subhalo mass.

Only the most massive subhalos (i.e., $V_{\rm max}/V_{200}>0.3$) deviate from this trend, and appear substantially less affected by tides than less massive subhalos. The median $V_{\rm max}/V_{\rm peak}$ is $ \sim 0.82$ for low mass subhalos, but climbs to $\sim 0.93$ at the massive end. Why do tides seem to affect more massive halos less? This is most likely a result of the rapid dynamical friction-driven evolution of massive halos, which tend to merge with the primary halo quickly after accretion. In other words, the (few) very massive halos present at any given time result from recent accretion events where tides have not had any substantive effect yet.

The results shown in  Fig.~\ref{fig:vmaxvpeak} can be used to statistically correct the distribution of $V_{\rm max}/V_{\rm 200}$ measured in cosmological simulations and to estimate the $V_{\rm peak}$ subhalo distribution of a given halo. The result is illustrated by the thick red line in  Fig.~\ref{fig:subhaloVF}, which shows the tide-corrected form of Eq.~\ref{eq:wang}. 
%As expected, this provides a good fit to the measured $V_{\rm peak}$ distribution of all satellites in APOSTLE primaries (shown by the dashed black(??) line). 
We shall hereafter adopt the tide-corrected version of  Eq.~\ref{eq:wang} (with Poisson scatter) to model the $V_{\rm peak}$ distribution of a halo of given $V_{200}$.
%??Should we give a fitting formula for this "tide-corrected" Vpeak distribution? It might be useful??

%%%%%%%%%%%%%%%%%%%%%%%%%%%%%%%%%%
\begin{figure*}
  \includegraphics[width=1.\linewidth]{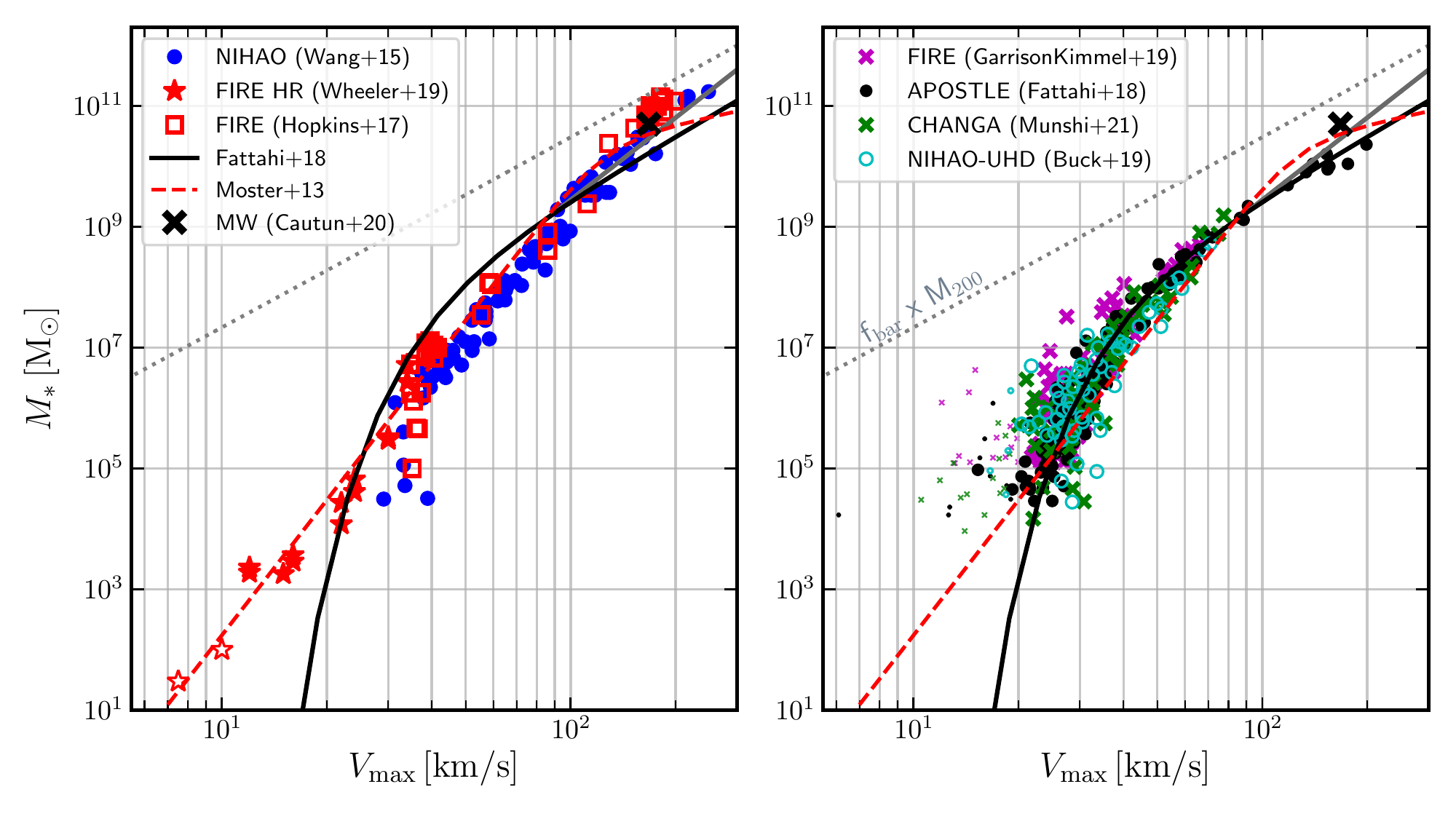}
  \caption{ $M_*$-$V_{\rm max}$ relations for simulated 'central' galaxies at $z=0$  from recent cosmological hydrodynamical simulations (see legends). These have been divided into two groups, depending on whether they either roughly follow a ``power-law''-like relation (left panel), or a relation with a sharp ``cutoff'' in $M_*$ at low $V_{\rm max}$ (right panel)
like the ones assumed in this work.
    The power-law results are well described by the extrapolated abundance-matching relation from \citet{Moster2013} (red dashed line); the cuttoff results may be approximated by the fit to APOSTLE data from \citet{Fattahi2018} (black line).
Open red star symbols indicate FIRE-HR galaxies with 1 and 15 stellar particles.  Backsplash galaxies in APOSTLE are shown as smaller black points.  In the right-hand panel, we mark all simulated galaxies from the other samples showing $V_{\rm max}<20$ km/s with smaller symbols as they are likely to have been affected by tides as well.
(Indeed, the NIHAO-UHD sample avoids most backsplash galaxies by selecting central dwarfs outside $2.5\times r_{200}$ of a massive primary, and shows in general $V_{\rm max}>20$ km/s.)
For reference, a dotted gray line marks the maximum total amount of baryons inside $M_{200}$ as expected from the cosmic mix, where $f_{\rm bar}=\Omega_{\rm bar}/\Omega_{\rm m}$.
  }
 \label{fig:mstarvmax}
\end{figure*}
%%%%%%%%%%%%%%%%%%%%%%%%%%%%%%%%%%

\subsection{The stellar mass-halo mass relation}
\label{SecM*Vpeak}

What is the stellar mass expected for a subhalo of given $V_{\rm peak}$? Fig.~\ref{fig:mstarvmax} motivates our choice of models for the stellar mass-halo mass relation. This figure shows the $M_*$-$V_{\rm max}$ relation reported for central galaxies at $z=0$
%(i.e., isolated and not tidally-stripped systems) 
selected from recent cosmological hydrodynamical simulations, as indicated in the legend. For central galaxies $V_{\rm peak}$ is in general achieved at $z=0$ (except for those centrals that have tidally-interacted in the past, which have been removed from our sample), and therefore the peak maximum circular velocity coincides with the maximum circular velocity at present-day, $V_{\rm max}$.  
When necessary, we have transformed quoted halo masses into $V_{\rm max}$ assuming they follow a Navarro-Frenk-White density profile \citep[hereafter, NFW,][]{Navarro1996,Navarro1997} with a mass-concentration relation as given by \citet{Ludlow2016}.

These simulations suggest two different behaviours for the $M_*$-$V_{\rm max}$ relation. On the left panel of Fig.~\ref{fig:mstarvmax} we have grouped simulations where $M_*$ and $V_{\rm max}$ seem better described by a simple power law that extends down from $V_{\rm max} \sim 200$ km/s to less than $\sim 10$ km/s, deep into the ultrafaint regime ($M_*\sim 10$-$10^2 M_\odot$). 

Interestingly, the power-law follows closely the extrapolated $M_*$-$V_{\rm max}$ relation from \citet{Moster2013}  (dashed red line),
\begin{equation}\label{eq:powerlaw}
M_*= {0.0702\,  \over [(m_1)^{-\beta} + (m_1)^{\gamma}]} \, M_{200}, 
\end{equation}
%M_*/M_\odot= M_{200}/M_\odot \times 2 \times N [(M_{200}/M_1)^{-\beta} + (M_{200}/M_1)^{\gamma}]^{-1}, 
%%M_*= {0.0702\, m_1 \over [(m_1)^{-\beta} + (m_1)^{\gamma}]} \, M_1, 
where $m_1=M_{200}/M_1$,  $M_1= 10^{11.59}\, M_\odot$, $\beta=1.376$, and $\gamma=0.608$.
As above, $M_{200}$ in this relation can be easily tranformed into $V_{\rm max}$ assuming an NFW density profile and a mass-concentration relation.
%??Should we explicitly quote the relation we have used??
%ISS this is the relation I use. I just transform the Vmax to M200 using NFW and M200(c) from Ludlow.

On the other hand, the panel on the right in Fig.~\ref{fig:mstarvmax} groups simulations whose results seem better described by a rapidly steepening relation between $M_*$ and $V_{\rm max}$ towards decreasing $V_{\rm max}$, suggesting the presence of a cutoff in the relation. Following  \citet{Fattahi2018},  this ``cutoff'' relation may be parametrized as: 
\begin{equation}\label{eq:cutoff}
M_*= \eta^\alpha \exp(-\eta^\mu) \, M_0 ,
\end{equation}
with 
%%$\eta=V_{\rm max}/54$ km/s, and $(M_0,\alpha,\mu)=(4\times10^8\, M_\odot,\,3.36,\,-2.2)$, 
$\eta=V_{\rm max}/50$ km/s, and $(M_0,\alpha,\mu)=(3\times10^8\, M_\odot,\,3.36,\,-2.4)$, 
shown by the solid black line in Fig.\ref{fig:mstarvmax}.

Although those authors fitted only results for central galaxies, an indistinguishable fit is obtained when adding the  $M_*$-$V_{\rm peak}$ data for APOSTLE satellites, which justifies the use of Eq.~\ref{eq:cutoff} to model the stellar content of a satellite of given $V_{\rm peak}$. (For centrals $V_{\rm peak}\approx V_{\rm max}$  at $z=0$, by construction.)
%
%We note that Eq.~\ref{eq:cutoff} is a slight modification of that originally proposed by \citet{Fattahi2018} to describe APOSTLE results (shown by a thin solid black line in the right panel of  Fig.~\ref{fig:mstarvmax}). This is because those authors fitted only results for central galaxies, whereas the APOSTLE data shown here includes the $M_*$-$V_{\rm peak}$ data for APOSTLE satellites as well. Interestingly, the changes are relatively minor, which justifies the use the $M_*$-$V_{\rm max}$ relation of centrals at $z=0$ to model the stellar content of a satellite of given $V_{\rm peak}$. (For centrals at $z=0$, $V_{\rm peak}\equiv V_{\rm max}$, by definition.)

The ``power-law'' and ``cutoff'' relations between stellar mass and peak velocity are the sole difference between the two models we explore in this paper. We emphasize that it is not our intention to categorize the different simulations into one or the other behaviour, but instead to motivate these two different analytical models that seem to describe well the current predictions from several simulations. 

Moreover, both models explored here are meant to be purely empirical, without being strongly linked to particular choices of subgrid physics or numerical resolution. The fact that most of current predictions from state-of-the-art numerical simulations align well with one of the two models, independent of their assumed galaxy formation physics and resolution, is reassuring and provides support to the approach presented in this work.

These two $M_*$-$V_{\rm peak}$ relations are plotted in both panels of Fig.~\ref{fig:mstarvmax} for ease of comparison (red dashed curve for ``power-law'' and solid black for ``cutoff''). The main differences between them are the behaviour at low $V_{\rm peak}$ and the slope of the relation at intermediate $V_{\rm peak}$, between $\sim30-80$ km/s. 
 Because their predictions differ for systems like the Milky Way, which we shall use to calibrate our models, we adopt a single $M_*$-$V_{\rm max}$ relation for $V_{\rm max}> 84$ km/s (where the \citealt{Fattahi2018} and the \citealt{Moster2013} lines cross each other). This is shown by the power-law solid gray line depicted in Fig.~\ref{fig:mstarvmax} and may be expressed as
\begin{equation}\label{eq:cutoff}
M_*/M_\odot=3.29\times 10^9 \, (V_{\rm max}/84\, {\rm km/s})^{4.52} ,
\end{equation}
applicable only for $V_{\rm max}>84$ km/s.

In what follows, we shall express the stellar mass-halo mass relation in terms of $V_{\rm peak}$, defined as the maximum circular velocity of a satellite before infall, or,  for centrals, as $V_{\rm max}$ at $z=0$.

%%%%%%%%%%%%%%%%%%%%%%%%%%%%%%%%%%
\begin{figure*}
\includegraphics[width=\linewidth]{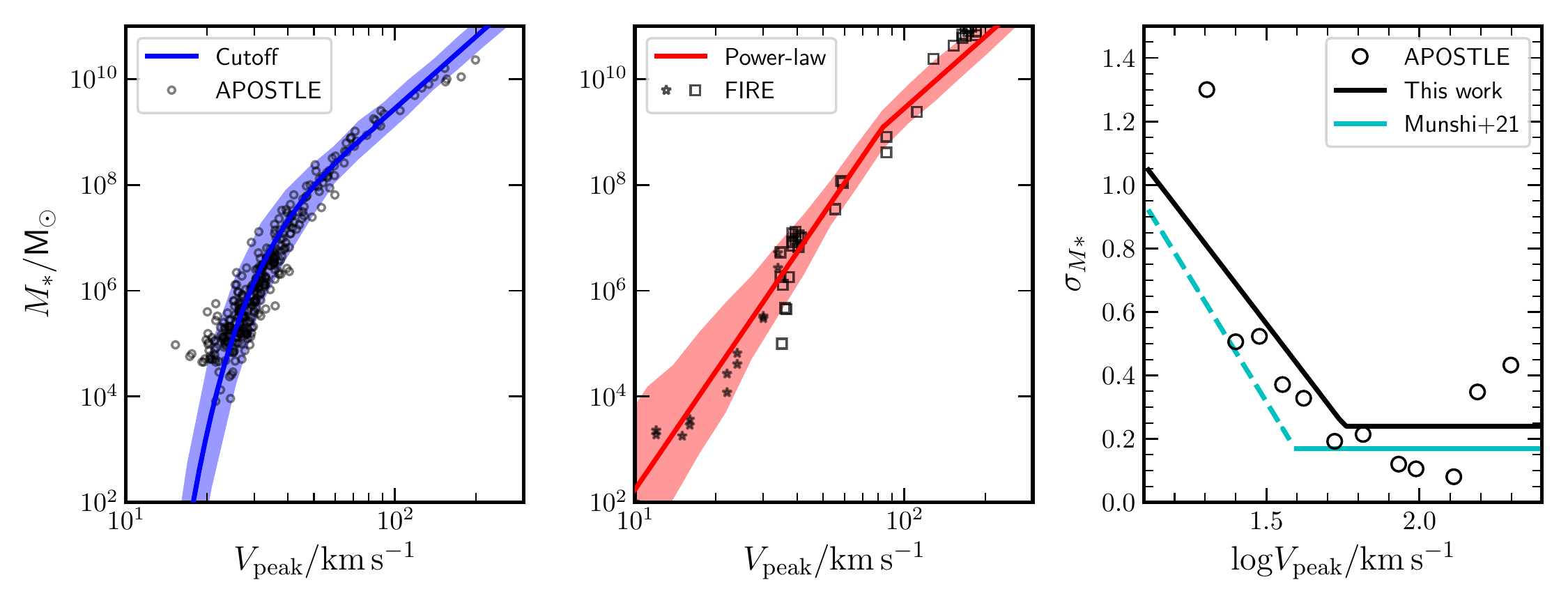}
\caption{ $M_*$-$V_{\rm peak}$ relations for the two models studied in this paper. Left: ``cutoff''; Middle: ``power-law''. The ``cutoff'' relation has been calibrated to match the APOSTLE simulation data for subhalo centrals at $z=0$ and for subhalo satellites of MW/M31 analog primaries before infall, when their circular velocities peak (black circles).
We show our ``power-law'' model in comparison to FIRE simulation data \citep[black symbols in middle panel;][]{Hopkins2018,Wheeler2019}.
The shaded bands show the $10-90$ percentile range in $M_*$ at given $V_{\rm peak}$.  The scatter is assumed to be a function of $V_{\rm peak}$, following Eq.~\ref{eq:scatter},
which is described in the rightmost panel.
Throughout this paper we assume the same scatter model for both the ``cutoff'' and ``power-law'' models. The black  line in the rightmost panel compares our assumed scatter (solid black line) to the scatter in $M_*$ measured in bins of $V_{\rm peak}$ for APOSTLE simulated galaxies (open circles) and to the model proposed by  \citet{Munshi2021} after transforming their $M_{\rm peak}$ parametrization to $V_{\rm peak}$ (see text for more details).
}
 \label{fig:msvpkrels}
\end{figure*}
%%%%%%%%%%%%%%%%%%%%%%%%%%%%%%%%%%

\subsubsection{Scatter}
\label{SecM*VpeakScatter}

As is clear from  Fig.~\ref{fig:mstarvmax}, the  
$M_*$-$V_{\rm peak}$ 
%stellar mass-halo mass
relation has substantial scatter. We account for this assuming that $M_*$ follows, at given $V_{\rm peak}$, a  log-normal distribution with a dispersion, $\sigma_{M_*}$, that increases toward decreasing halo masses. Following \citet{GarrisonKimmel2017} and  \citet{Munshi2021}, we parametrize $\sigma_{M_*}$ as a broken power-law of $V_{\rm peak}$, as illustrated by the solid black line in the right-hand  panel of Fig.~\ref{fig:msvpkrels},
%\begin{equation}\label{eq:scatter}
%\sigma=\sigma_0 + \kappa \, (\log_{10} M_{200}/M_2),
%\end{equation}
%with $\sigma_0=0.24$ dex, $\kappa=-0.4$, and $M_2=3\times10^{10}\, M_\odot$.  
\begin{equation}\label{eq:scatter}
\sigma_{M_*}=\begin{cases} \sigma_0 & V_{\rm peak}>57\, \text{km/s}\\
 \kappa \log_{10} (V_{\rm peak}/V_0) & V_{\rm peak}<57\, \text{km/s}
\end{cases}
\end{equation}
with $\sigma_0=0.24$ dex, $\kappa=-1.26$, and $V_0=88.6\,\text{km/s}$.
%For comparison, we show the scatter measured for APOSTLE galaxies (see open black circles in the right panel of Fig.~\ref{fig:msvpkrels}).
These parameters have been chosen arbitrarily but loosely guided by the measured scatter in APOSTLE galaxies (see open black circles in the right panel of Fig.~\ref{fig:msvpkrels}) and by the scatter parametrization to CHANGA galaxies in \citet{Munshi2021} (see cyan line). Note that \citet{Munshi2021} parametrizes the scatter in terms of $M_{\rm peak}$ instead of $V_{\rm peak}$, 
but indicate that in the latter case they find a scatter floor of 0.17 dex and an increasing scatter that reaches $\sim 1$ dex at their lowest $V_{\rm peak}$s.  As an approximation, here we assume it reaches the same scatter at low $V_{\rm peak}$  as it does with their $M_{\rm peak}$ based model (dashed line).

%The left and middle panels of Fig.~\ref{fig:msvpkrels} show the  assumed ``cutoff"  and ``power-law" relations, respectively, where the cutoff relation is compared to the APOSTLE results used to motivate Eq.~\ref{eq:scatter}.  The shaded bands delineate the 10-90 percentile scatter in $M_*$ at given $V_{\rm peak}$. As expected, the assumed cutoff relation reproduces the APOSTLE results well. Assuming the same scatter, the ``power-law'' model also reproduces the results of the FIRE simulations well, as shown in the middle panel of Fig.~\ref{fig:msvpkrels}. 
For simplicity, we assume that  both the ``cutoff" and the ``power-law'' model have the same $M_*$ scatter dependence on $V_{\rm peak}$ given by Eq.~\ref{eq:scatter}. The shaded bands in the left and middle panels of Fig.~\ref{fig:msvpkrels} indicate the resulting $10$-$90$ percentiles in the $M_*$ distribution at a given $V_{\rm peak}$ assuming Eq.\ref{eq:scatter} in each model.  
The assumed cutoff relation with scatter reproduces the APOSTLE results well (see left panel of Fig.~\ref{fig:msvpkrels}). The middle panel of Fig.~\ref{fig:msvpkrels} shows that our choice, albeit arbitrary, also accommodates well other simulations too, such as FIRE, which we take as further validation of our assumed scatter model. Note that, while at fixed $V_{\rm peak}$ the scatter in $M_*$ is identical in both models, the ``cutoff"  model is steeper than the ``power-law" model at low $V_{\rm peak}$,  and therefore the shaded area is approximately constant in contrast with the visible increase in dispersion seen in the middle panel for the ``power-law".

We have checked that changing the details of this scatter model (i.e.  scatter floor, steepness of slope) makes negligible difference to results with the ``cutoff" model,  because changes apply to the $V_{\rm peak}$ regime below the intrinsic  threshold.  Although changes do affect somewhat results with the  ``power-law'' model,  the final relative differences between the satellite mass functions obtained with the ``cutoff" and ``power-law'' models  remain robust.

\subsection{Stellar mass loss}
\begin{figure}
\includegraphics[width=\linewidth]{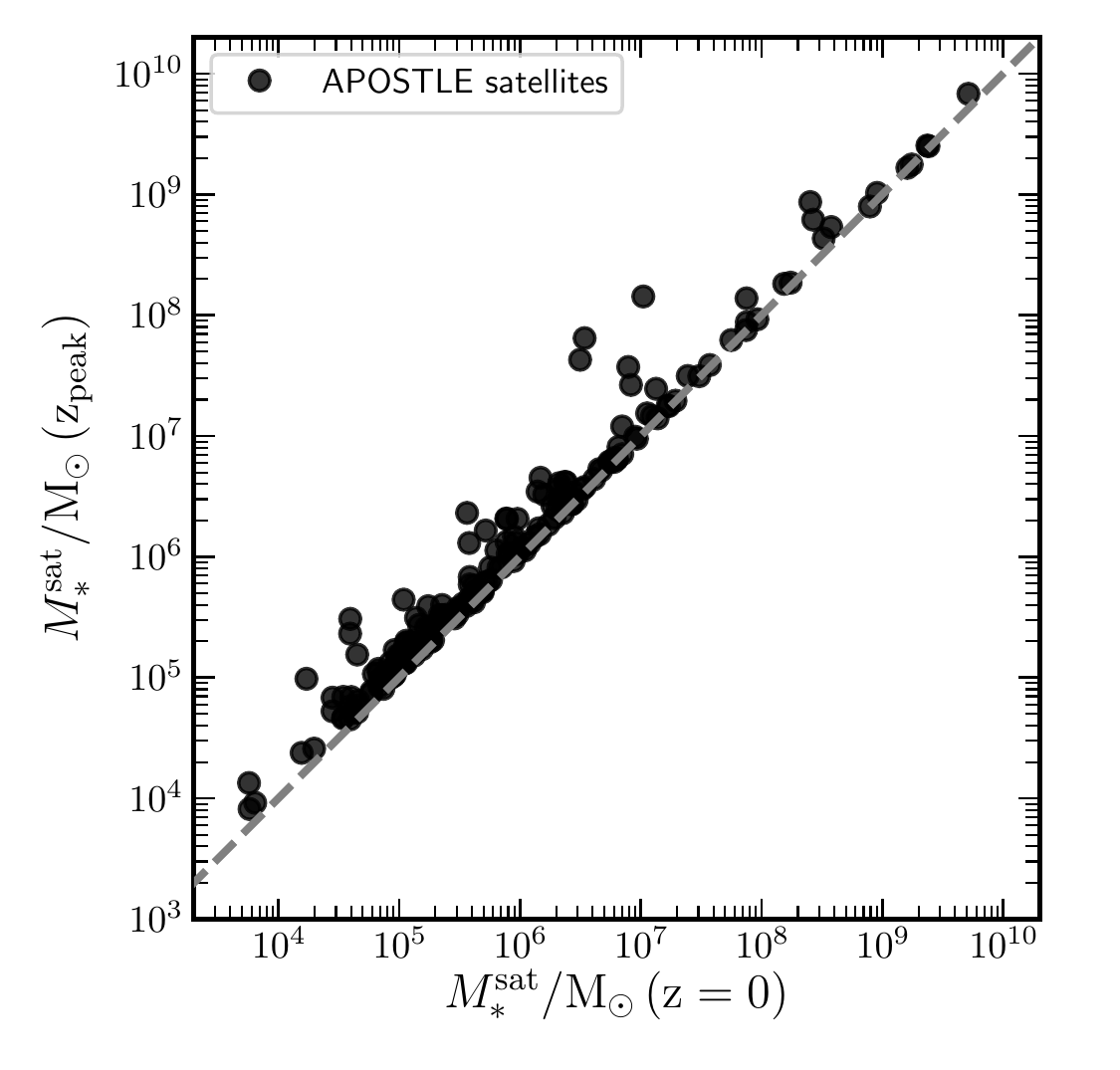}
\caption{ Stellar mass  of APOSTLE satellite galaxies measured at $z=0$ versus that measured at the time when the maximum circular velocity  of a subhalo peaks, $z_{\rm peak}$.  
The 1:1 correspondence is marked with a dashed gray line.
}
 \label{fig:msz0zpk}
\end{figure}

As mentioned above, our simple modeling shall neglect tidally induced stellar mass loss in subhalos. This is clearly a simplification, but finds support in the results for APOSTLE satellites, which show that the effects of stellar mass loss are quite modest. This is shown in Fig.\ref{fig:msz0zpk}, where we plot the stellar mass of APOSTLE satellite galaxies at $z=0$ versus that at $z_{\rm peak}$, the redshift when its maximum circular velocity peaked. Unlike $V_{\rm max}$, $M_*$ changes, on average, very little after infall into the main halo. Half of APOSTLE satellites have lost less than $\sim 22\%$ of their peak mass, and only $10\%$ have lost more than $\sim58\%$ since infall. In the interest of simplicity, we have decided not to include any corrections for stellar mass loss, but have checked that none of our main conclusions are altered if a correction of the magnitude suggested by Fig.~\ref{fig:msz0zpk}, is implemented.

\subsection{The cutoff and ``power-law'' models}

The assumptions discussed above allow us to compute the expected satellite stellar mass function for a system of arbitrary virial mass. To summarize, for a halo of given $V_{200}$ we first draw a realization of the subhalo $V_{\rm peak}$ function consistent with the tide-corrected Eq.~\ref{eq:wang}, assuming Poisson scatter. 
For each subhalo, we then draw a stellar mass using either the ``cutoff'' or the ``power-law'' models described in Sec.~\ref{SecM*Vpeak}, with scatter as given by Eq.~\ref{eq:scatter}.  Unless otherwise specified, we shall always show median results obtained by combining at least $\sim 100$ independent realizations of each primary, together with the 10-90 percentile range.
We have confirmed that this number of realizations yields converged results by running our model with up to $\sim 5$ times more iterations with which we find no significant differences.

%%%%%%%%%%%%%%%%%%%%%%%%%%%%%%%%%
\begin{figure*}
\includegraphics[width=\linewidth]{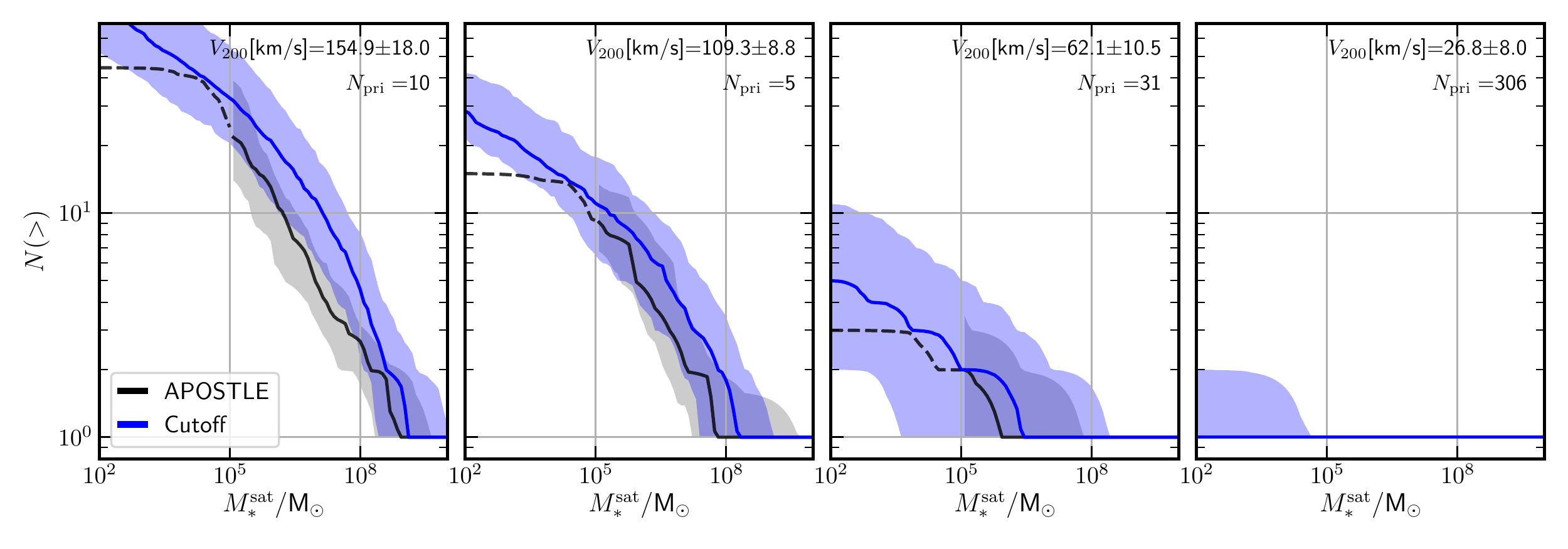}
\caption{Satellite stellar mass functions of APOSTLE primaries (grey) compared to those predicted by the ``cutoff'' model (blue).
Each panel corresponds to results for different bins in virial velocity of the primaries, $V_{200}$, as given in the legend. The number of APOSTLE primaries in each bin, $N_{\rm pri}$,  is also indicated in the legend. 
%The models assume the same $V_{200}$s  as those of APOSTLE primaries in each bin. 
The blue lines show results from the ``cutoff'' model as described in Sec.~\ref{sec:model}, applied to primaries with the same $V_{200}$ as APOSTLE ones in each bin. 
Solid  lines show the median results and shaded areas the 10-90 percentile range. 
For APOSTLE,  results below $M_*<10^5$ M$_\odot$ are shown with a dashed linestyle to indicate that objects below this mass are likely not well resolved.
}
\label{fig:apostle_model}
\end{figure*}
%%%%%%%%%%%%%%%%%%%%%%%%%%%%%%%%%

\section{Results}\label{sec:results}

%%%%%%%%%%%%%%%%%%%%%%%%%%%%%%%%%%%%%%%%%%%%%%%%%%%%%%%%%%%%%%%%
\begin{figure*}
\includegraphics[width=\linewidth]{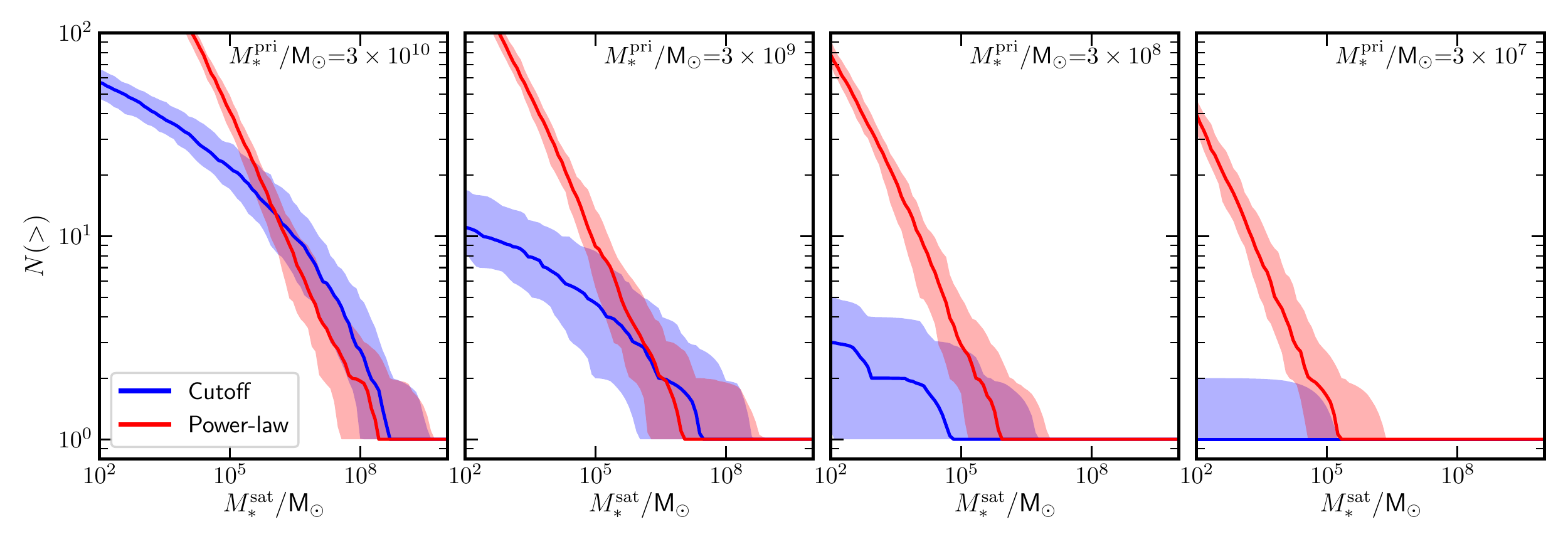}
\caption{ Comparison of the satellite stellar mass functions predicted by the ``cutoff'' (blue) and ``power-law'' (red) models. Each panel shows results assuming a different fixed value of  the stellar mass of the primary, $M_*^{\rm pri}$ (see legend).
 Solid lines show median results and shaded bands the 10-90 percentile range.
}
 \label{fig:models_smf}
\end{figure*}
%%%%%%%%%%%%%%%%%%%%%%%%%%%%%%%%%%%%%%%%%%%%%%%%%%%%%%%%%%%%%%%%
\subsection{The cutoff model and APOSTLE}\label{ssec:compareAP}

We start by comparing the results of the ``cutoff'' model with satellite mass functions from the APOSTLE simulations. 
We do this to check that our ``cutoff'' model is able to roughly reproduce the APOSTLE satellite stellar mass functions down to the resolution allowed by the simulations. Indeed, while we have chosen an average ``cutoff'' $M_*$-$V_{\rm peak}$ relation based on APOSTLE, it is not obvious a priori that our simple model can yield satellite mass functions overall consistent  with APOSTLE results. 

For example, our analytical ``cutoff" model includes a fully independent sampling of the subhalo mass function directly taken from $\Lambda$CDM simulations and corrected statistically by tidal stripping, and does not use the subhalo mass functions from APOSTLE. A good agreement between our analytical model and the APOSTLE results is a necessary benchmark for our analytical models.

This comparison is shown in Fig.~\ref{fig:apostle_model}, where each panel shows, in grey, the APOSTLE satellite mass functions for central galaxies, binned by halo virial velocity. The average $V_{200}$ and standard deviation in each bin is given in the legend of each panel. Solid lines show the median satellite mass function in the bin, while the shaded area represents the 10-90 percentile distribution.
 
Although we show mass functions down to stellar masses as low as $M_*>10^2\, M_\odot$ we note that objects  with $M_*<10^5\, M_\odot$ in APOSTLE  are resolved with fewer than 10 star particles. Therefore, below that mass APOSTLE results are best regarded as lower limits rather than actual simulation predictions.

By construction, the first bin (leftmost panel) includes the 10 primaries that are considered MW and M31 analogs in the APOSTLE volumes. For these APOSTLE primaries ($V_{200}\approx 150$ km/s, or, equivalently, $M_{200}\sim 1.2\times 10^{12}\, M_\odot$), the median number of satellites with
$M_*>10^5$ M$_\odot$  is $\sim 24.1^{+18.5}_{-6.7}$, where the uncertainties represent the 10-90 percentile range. 
  
For dwarf primaries with $V_{200}\approx 62$ km/s (third panel from the left) the number of satellites in APOSTLE is drastically reduced by the cutoff in the $M_{*}$-$V_{\rm peak}$ relation, with a median of only $\sim 2.0^{+1.7}_{-1.0}$ satellites with $M_*>10^5$ M$_\odot$.
Finally, the last panel shows that no luminous satellites are found in APOSTLE around primaries with $V_{200}\lesssim 35$ km/s.

%%%%%%%%%%%%%%%%%%%%%%%%%%%%%%%%%%%%%%%%%%%%%%%%%%%%%%%%%%%%%%%%
\begin{figure*}
\includegraphics[width=0.8\linewidth]{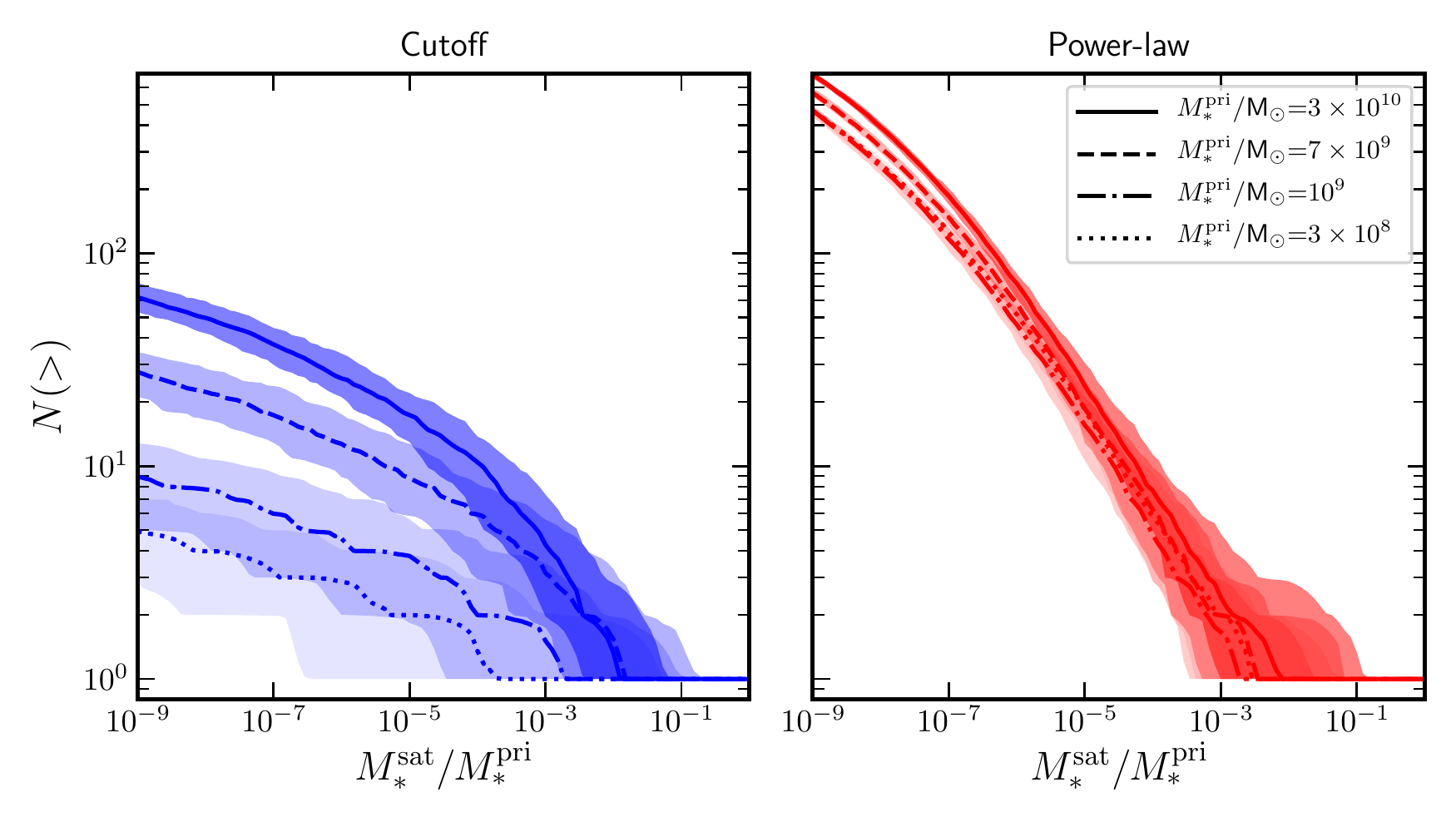}
\caption{ \textit{Scaled} satellite stellar mass functions (i.e., $M_*^{\rm sat}/M_*^{\rm pri}$) predicted by the ``cutoff'' (blue) and ``power-law'' (red) models. Different linestyles show results assuming  a different fixed value of the stellar mass of the primary, $M_*^{\rm pri}$ (see legend).
Lines show median results and shaded bands the 10-90 percentile range.
 Note that for $M_*^{\rm pri}\lesssim 5\times10^{9}\,M_\odot$ the ``power-law'' mass function becomes independent of primary mass, as discussed by \citet{Sales2013}.
}
 \label{fig:msat_norm}
\end{figure*}
%%%%%%%%%%%%%%%%%%%%%%%%%%%%%%%%%%%%%%%%%%%%%%%%%%%%%%%%%%%%%%%%

The blue bands in Fig.~\ref{fig:apostle_model} show the results of the ``cutoff'' model, applied to a sample of primaries whose number and $V_{200}$ distribution matches that in each APOSTLE bin. 
%\bt{This is, for each $V_{200}$ we run the complete model as described in Sec.~\ref{sec:model}, where there is no  assumption based on the APOSTLE simulations except for the average $M_*$-$V_{\rm peak}$ relation.\footnote{\bt{We emphasize that we are following the model described in Sec.~\ref{sec:model}, based on sampling the \citet{Wang2012} subhalo mass function including a  tidal-stripping correction, and that we are not populating APOSTLE's actual satellite mass functions. }}
We use $10$ independent realizations of  the satellite mass function of each primary to obtain robust results.

There is in general good agreement between the analytical ``cutoff'' model and the  simulation results, especially for satellites with $M_*>10^5\, M_\odot$.  Even the number of massive ($M_*>10^8$ M$_\odot$) satellites is well reproduced, with a median of $\sim 1$-$2$ LMC or SMC-mass satellites expected around MW-mass primaries (leftmost panel).

This is not unexpected, given that we have motivated the model on APOSTLE results, but it provides validation for our approach. It also  allows us to predict the population of dwarfs fainter than currently resolved by APOSTLE and other simulations. Importantly, the ``cutoff'' model predicts a steady decline in the number of satellites surrounding dwarfs of decreasing mass, approaching zero as the mass of the primary approaches the threshold mass (rightmost bin in Fig.~\ref{fig:apostle_model}).

\subsection{Cutoff vs power-law model satellite mass functions}\label{ssec:compare}

We now compare the satellite stellar mass functions predicted by each model, as a function of the stellar mass of the primary. This is shown in Fig.~\ref{fig:models_smf}, where each panel corresponds to a different $M_*^{\rm pri}$, given in the panel legends  (cutoff in blue, power-law in red). The most obvious difference is the large difference in the number of faint satellites predicted by each model. Hundreds of ultrafaints with $M_*> 10^2\, M_\odot$ are expected in the ``power-law'' model, even for primaries as faint as the Magellanic Clouds, whereas ultrafaint  numbers are much less numerous in the case of the ``cutoff'' model.

The difference between models is more clearly appreciated  when comparing the normalized satellite mass functions; i.e., the satellite mass function expressed in terms of  $M_*^{\rm sat}/M_*^{\rm pri}$. This is shown in Fig.~\ref{fig:msat_norm} for all primary stellar mass bins in the ``cutoff'' model (left) and ``power-law'' model (right). For the ``power-law'' model the normalized satellite mass function changes little with primary stellar mass. In particular, primaries with $M_*^{\rm pri}<10^9\; \rm M_\odot$ would be expected to share the same normalized satellite mass function, as shown by the overlap of the red dotted and long-dash-dotted lines in the right-hand panel of Fig.~\ref{fig:msat_norm}. 

As discussed by \citet{Sales2013}, this near ``self similarity'' arises because the subhalo mass function and the stellar mass-halo mass relation in this model are both close to power-laws, and thus scale-free. This is particularly true at $M_* < 10^9 \; \rm M_\odot$ (see middle panel Fig.~\ref{fig:msvpkrels}), which explains why lower mass bins overlap in their normalized satellite mass function. On the other hand, if the stellar-halo mass relation is not scale-free, as is the case for the ``cutoff'' model, the normalized satellite mass function declines with decreasing primary mass (see blue curves on the left panel of Fig.~\ref{fig:msat_norm}). The large differences between models suggest that the satellite mass function around isolated primaries spanning a wide range of mass (and, in particular, including dwarfs) may be used to infer the shape of the stellar mass-halo mass relation at the faint end.

%%%%%%%%%%%%%%%%%%%%%%%%%%%%%%%%%%%%%%%%%%%%%%%%%%%%%%%%%%%%%%%%
\begin{figure*}
\includegraphics[width=0.8\linewidth]{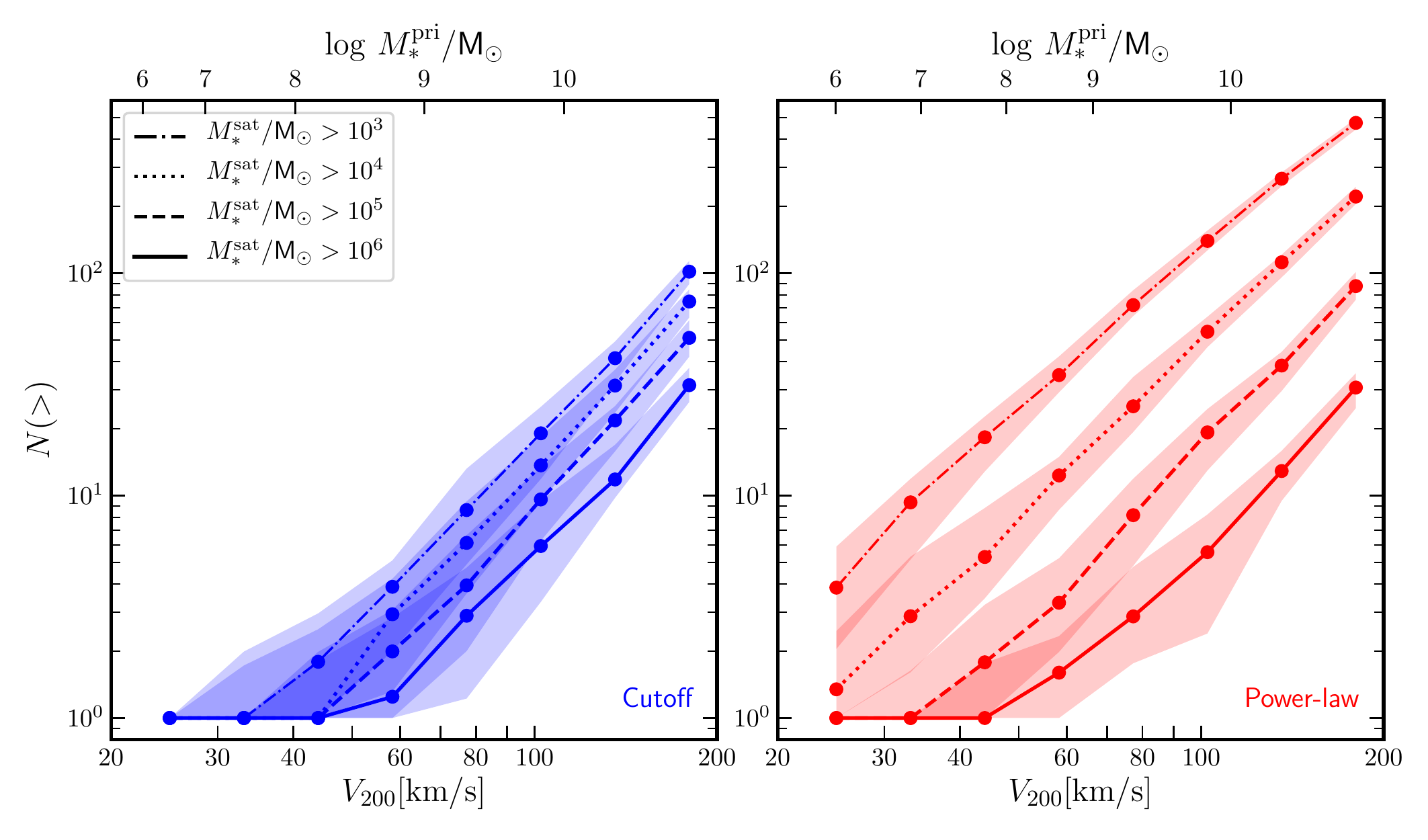}
\caption{The cumulative number of satellites with $M_*^{\rm sat}$ above certain values (see legend) predicted by the ``cutoff'' and ``power-law'' models for primaries with fixed values of $V_{200}$, as indicated in the x-axis. The corresponding $M_*^{\rm pri}$ values, according to each model, can be read from the top x-axis.
 Lines show median results and shaded bands show the 10-90 percentile range.
}
\label{fig:numsat_v200}
\end{figure*}
%%%%%%%%%%%%%%%%%%%%%%%%%%%%%%%%%%%%%%%%%%%%%%%%%%%%%%%%%%%%%%%%

Another, perhaps more intuitive contrast between models may be obtained comparing the expected total number of satellites more massive than a given stellar mass as a function of primary halo virial mass. We show this in Fig.~\ref{fig:numsat_v200}, where different line-styles indicate the cumulative number of satellites above a given $M_*^{\rm sat}$, as labeled on the left panel, as a function of either the virial velocity of the host (a proxy for the primary halo mass; lower x-axis) or the corresponding stellar mass of the primary according to each of the two models (upper x-axis).

For massive satellites (i.e., $M_*^{\rm sat}>10^6 \; \rm M_\odot$, solid line) the predictions of the two models are rather similar, with $\sim 2.5$ satellites on average in hosts with $V_{200} = 75$ km/s and $10$-$30$ satellites for hosts in our most massive bin, $V_{200} \geq 150$ km/s. However, the predictions of the two models differ appreciably when considering fainter satellites and, in particular, in the regime of ultrafaint dwarfs.

For example, in the ``power-law'' model, a dwarf primary with $M_*^{\rm pri} \sim 10^9 \; \rm M_\odot$ (like the LMC) is expected to host $\sim 70$ satellites with $M_*^{\rm sat} \geq 10^3\; \rm M_\odot$, $\sim 50$ of which would be ultrafaint ($M_*^{\rm sat} < 10^5\, M_\odot$). On the other hand, in the ``cutoff'' model only $8$ satellites are expected with $M_*^{\rm sat} \geq 10^3\, M_\odot$ for the same primary. As we have seen before, the population of ultrafaint satellite dwarfs is heavily suppressed in models with a sharp cutoff in the stellar mass-halo mass relation like the one explored here. Deep imaging and spectroscopic surveys of the surroundings of isolated dwarfs designed to constrain the satellite population within their virial radius should thus yield key insights into the stellar mass-halo mass relation at the faint end.

\section{Models vs. Observations}
\label{ssec:data}

\subsection{Milky Way and M31 satellites}
\begin{figure*}
\includegraphics[width=\linewidth]{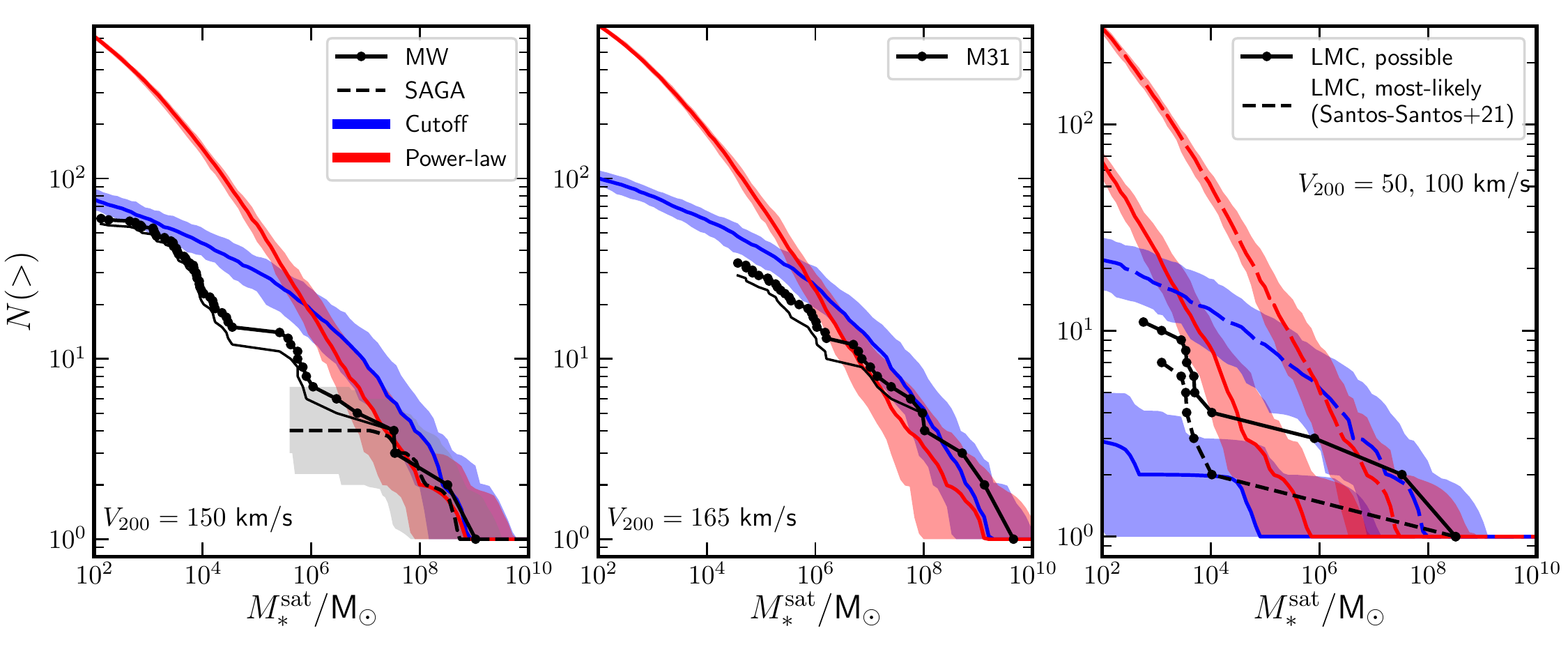}
\caption{ Predictions of the ``cutoff'' (blue) and ``power-law'' (red) models for the satellite stellar mass functions of observed host galaxies. Left: Milky Way; Middle: M31; Right: LMC (note the different y-axis limits). For each case, we assume a fixed value of $V_{200}$ of the primary motivated by literature estimates (see text for details).
 Solid lines show median results and shaded bands the 10-90 percentile range.
 Black points show the observed satellite stellar mass function of each host.  In the case of MW and M31, thicker lines show results for satellites within 300 kpc, while thinner lines correspond to satellites within $r_{200}$ (214 kpc for $V_{200}=150$ km/s, and 236 kpc for $V_{200}=165$ km/s, assuming h=0.7).
The first panel additionally shows the satellite stellar mass function of MW-mass analogs observed as part of the SAGA survey \citep{Mao2020}.
For the case of the LMC, we show  model results assuming $V_{200}=50$ (colored, solid) as well as $100$ (colored, long-dashed) km/s, compared to likely Magellanic satellites according to \citet{Santos-Santos2021}.
}
\label{fig:modelvsdata}
\end{figure*}

The most complete available census of faint satellites is in the Local Group, which provides therefore a good testbed for the ideas explored above. We compare in Fig.~\ref{fig:modelvsdata} the predictions of the theoretical models with data for MW and M31 satellites (left and middle panels)\footnote{Stellar masses for observed Local group satellites have been estimated using luminosities from \citet{McConnachie2012} and assuming appropriate mass-to-light ratios according to \citet{Woo2008}.}.  Black symbols connected by a solid curve show the observational data, taken from \citet{McConnachie2012}'s updated compilation of Local Group dwarfs where objects within $300$ kpc of the MW/M31 are considered satellites.

For the models, we choose a virial velocity of $V_{200}=150$ km/s for the MW, and a somewhat larger $V_{200}=165$ km/s for M31, in agreement with current available mass constraints \citep[see; e.g.,][]{Cautun2020,Sofue2015,Fardal2013}. Each virial velocity is sampled $100$ times; the resulting median and $10$th-$90$th percentiles are shown in blue for the ``cutoff'' model and in red for the ``power-law'' model.

%%%%%%%%%%%%%%%%%%%%%%%%%%%%%%%%%%%%%%%%%%%%%%%%%%%%%%%%%%%%%%%%
%%%%%%%%%%%%%%%%%%%%%%%%%%%%%%%%%%%%%%%%%%%%%%%%%%%%%%%%%%%%%%%%

The number of MW satellites  with $M_*^{\rm sat} \geq 10^6\; \rm M_\odot$ is in reasonable agreement with both models (see left panel in Fig.~\ref{fig:modelvsdata}), as well as with data from the SAGA survey, which targeted the bright end of the satellite population within $300$ kpc of MW-like primaries \citep{Mao2020}. We note that our models refer to satellites within the virial radius of the assumed halo ($r_{200}\sim 215$ kpc for our choice of $V_{200}=150$ km/s) rather than the $300$ kpc used in the observational data. The thin black line in the left-hand panel of Fig.~\ref{fig:modelvsdata} shows the known MW satellites inside that smaller radius; the difference is quite small.

Both the ``power-law'' and the ``cutoff'' model predict the same number of satellites with $M_*>10^6\, M_\odot$ (roughly $\sim 18$), interestingly well in excess of the known number of such systems orbiting the Milky Way.  The discrepancy worsens between $10^4$ and $10^6\, M_\odot$, where the MW satellite mass function appears to have a sizable ``gap''. It is unclear what the significance of such gap may be, but it is tempting to associate it with increasing incompleteness in observational detections \citep[see; e.g., the discussion in][]{Fattahi2020}. The numbers climb rapidly in the $10^2$-$10^4\, M_\odot$ range, to almost match the predictions of the ``cutoff'' model.

As discussed in Sec.~\ref{ssec:compare}, it is in the ultrafaint regime where the ``power-law'' and ``cutoff'' models can be best differentiated. For faint dwarfs with $M_* ^{\rm sat} < 10^5\; \rm M_\odot$, the ``cutoff'' model predicts substantially fewer ultrafaints than the ``power-law'' model. Interestingly, this comparison suggests that  if the stellar mass-halo mass relation does indeed have a low-mass cutoff, the majority of ultrafaint dwarfs in the MW might have already been discovered, leaving little room to accommodate a large missing population of ultrafaints. On the other hand, the ``power-law'' model suggests the presence of a numerous, yet undetected population of ultrafaints in the MW halo. Upcoming surveys of the MW satellite population, especially those which account for satellites hidden behind the disk, or missing due to the incomplete spatial and surface brightness coverage of existing surveys, should be able to distinguish clearly between the two models proposed here.

%This agrees with recent semianalytic models, which argues that the MW ultrafaint population may have been further enhanced by the Gaia-Enceladus merger \citep{Bose2020} and by the relatively recent accretion of the LMC \citep{Nadler2020}.  Taken at face value, however, current data in the MW therefore seem to disfavor a simple power-law behaviour for the faint end of the stellar mass-halo mass relation.

%\subsection{M31 satellites}
The middle panel of Fig.~\ref{fig:modelvsdata} compares model predictions with current estimates of the M31 satellite population. Although the surveyed population in M31 does not go as deep as in the MW, the total number of satellites with $M_* \geq 10^5\; \rm M_\odot$ seems to fall below the ``power-law'' model predictions, at least for the virial mass explored here.  There is a hint that the observed satellite mass function compares more favourably with the ``cutoff'' model, which predicts roughly half as many satellites in that mass range as the ``power-law'' model.

The ``cutoff'' model predicts at least $\sim 50$ new ultrafaint M31 satellites in the range $10^2 < M_*/\rm M_\odot < 10^{4.6}$ (the mass of And XX, the least massive M31 satellite known), bringing the total population to $\sim 90$-$110$ total dwarfs above a stellar mass $100\; \rm M_\odot$. By contrast, the ``power-law'' model predicts a total of $\sim 680$-$740$ satellites with $M_*>10^2\, M_\odot$. We note that these numbers are quite sensitive to the choice of virial mass for the M31 halo; doubling the mass (i.e., increasing $V_{200}$ to $208$ km/s) would yield roughly twice as many satellites for either model, although the relative differences in mass function shape would be preserved.

\subsection{Satellites of isolated LMC-like dwarfs}

%Within the Local Group, the LMC offers the best chances to constraining the satellite population of dwarf primaries. This is not only because of its relatively high stellar mass, being one of the most massive dwarfs in the Local Group, but also because its recent infall less than 2 Gyrs ago allows for a phase-space reconstruction of the dwarfs that were once members of its ``Magellanic system" before accretion into the MW \citep[see ][]{Sales2011, Kallivayalil2018}.

Finally, the right-hand panel of Fig.~\ref{fig:modelvsdata} shows the predictions for the satellite population of isolated dwarf galaxies with stellar mass comparable to that of the LMC ($M_*\sim 3\times 10^9 \rm M_\odot$) , or, more precisely, dwarf primaries inhabiting halos with  $V_{200}$ in the range $50$ to $100$ km/s. This is consistent with the virial mass range ($2.5<M_{\rm 200}/10^{10}\, M_\odot< 45$) of galaxies with comparable stellar mass in the APOSTLE simulations \citep[see; e.g.,][]{Santos-Santos2021}. These authors use kinematic information to identify LMC-associated
dwarfs; their list of most likely LMC satellites include 7 satellites: the SMC, Hydrus 1,  Horologium 1,  Carina 3,  Tucana 4, Reticulum 2,  and Phoenix 2 (dashed black line). A less likely, but still plausible association is also ascribed to  Carina,  Horologium 2,  Grus 2, and  Fornax, bringing the total to 11 (solid black line).

In the context of the ``cutoff'' model, these numbers seem to rule out a virial velocity as low as $50$ km/s (bottom blue curve), and suggest a virial velocity a little below $100$ km/s (top blue curve). In contrast, a virial velocity near the lower bound would be favoured in the case of the ``power-law'' model. An LMC halo as massive as $100$ km/s seem quite inconsistent with the data in this case. Note that the predictions of the two different models differ substantially even for satellites with $M_* \sim 10^4 \; \rm M_\odot$. This limit seems within reach of what may be achievable in future surveys of LMC-like primaries, turning them into strong constraints of the stellar mass-halo mass relation of faint galaxies, a subject we address in more detail below.

%%%%%%%%%%%%%%%%%%%%%%%%%%%%%%%%%%%%%%%%%%%%%%%%%%%%%%%%%%%%%%%%
\begin{figure}
\includegraphics[width=\linewidth]{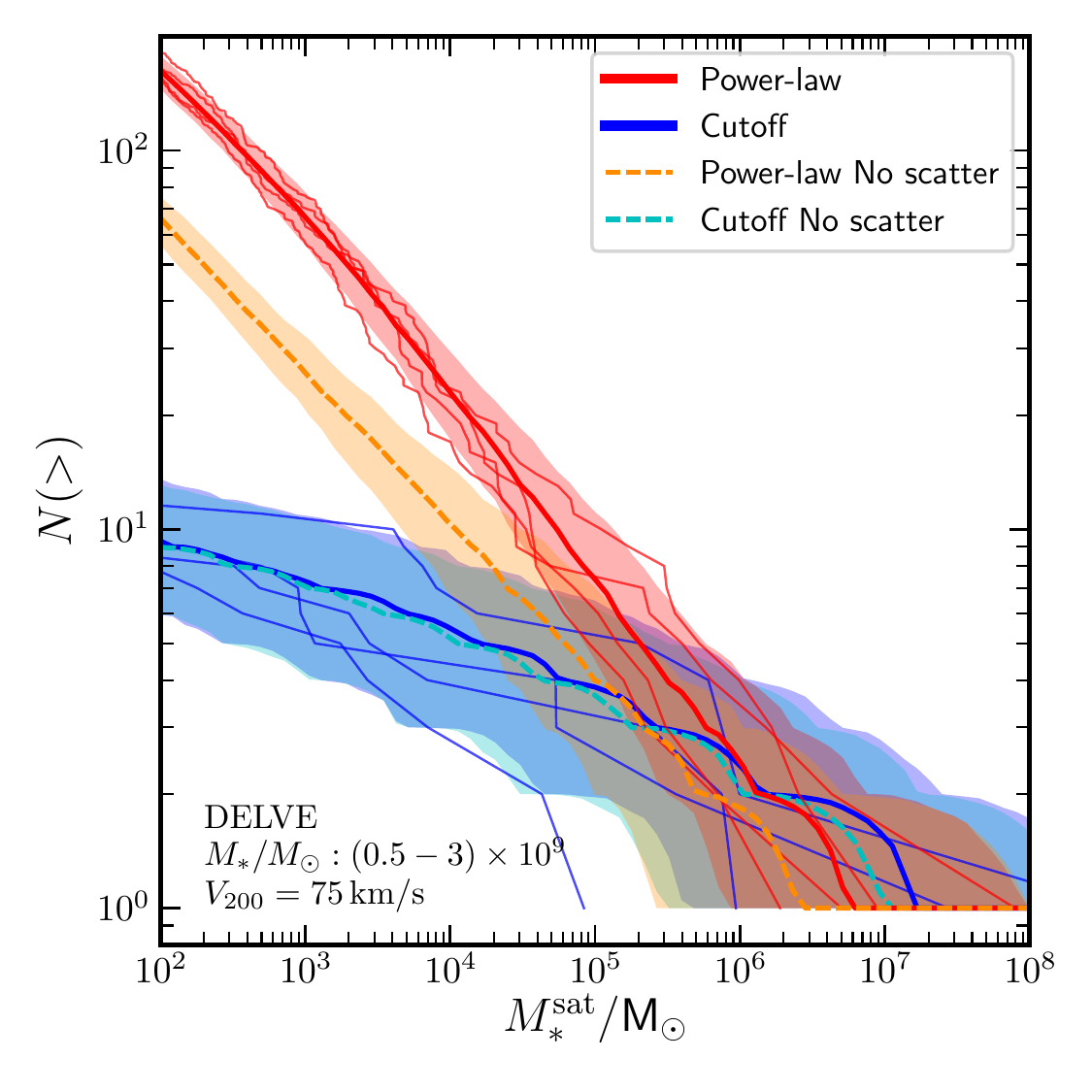}
\caption{
  Satellite stellar mass functions predicted by the ``cutoff'' (blue) and ``power-law'' (red) models, assuming a fixed primary halo virial velocity of $V_{200}=75$ km/s.  A solid line shows the median result, while the shaded band shows the 10-90 percentile range.  In addition, the dashed lines show the results obtained assuming \textit{no scatter} in the $M_*$-$V_{\rm peak}$ relation. This has a strong effect on the power-law results (in orange), but affects very little the cutoff ones.  These results summarize the predictions of our two models for the satellite mass functions of the 4 DELVE Magellanic analogs NGC300, NGC55, IC5152 and SextansB, with stellar masses in the $[0.5,3]\times 10^9\, M_\odot$ range \citep{Drlica2021}.  Thin individual lines show the results for 4 individual random realizations of the model (with scatter), to illustrate the expected object-to-object variation.  }
\label{fig:modelfordelve}
\end{figure}
%%%%%%%%%%%%%%%%%%%%%%%%%%%%%%%%%%%%%%%%%%%%%%%%%%%%%%%%%%%%%%%%

\subsection{Predictions for future surveys}
\label{ssec:data_future}

Beyond the Local Group, several ongoing (and future) observational efforts have the potential to measure the satellite population of isolated LMC-like galaxies, and thus deliver strong constraints on the stellar mass-halo mass relation at the faint end.  To reduce fluctuations due to object-to-object scatter, it is desirable to survey several primaries of similar stellar mass while simultaneously reaching the ultrafaint satellite regime.  This is why dwarf galaxies are the most promising primaries: within the Local Volume (i.e., within $10$ Mpc from the Milky Way) there are only $8$ MW-like galaxies ($M_* \geq 10^{10.5}\, M_\odot$) outside the Local Group but there are $112$ known dwarfs with $10^8 < M_*/\rm M_\odot < 10^{9.5}$ \citep{Tully2009,Tully2016}. %Of these, the more massive LMC-analogs should host the most numerous ultrafaint satellite population.

As an example, we provide in Fig.~\ref{fig:modelfordelve} expectations from the ``cutoff'' and ``power-law'' models for the satellite mass function of $4$ LMC-like dwarfs expected to be surveyed as part of the DES DELVE campaign \citep{Drlica2021}. This includes NGC~300, NGC~55, IC~5152 and Sextans~B, which span a stellar mass range $M_*=[0.5,3] \times 10^9\; \rm M_\odot$.
The model predictions are based on $100$ realizations of dwarfs with fixed virial velocity, $V_{200}=75$ km/s, and are shown by the top red curves and the bottom blue curves. (Thin lines correspond to $4$ individual realizations, to illustrate the expected object-to-object scatter.)

As in our earlier discussion, this figure makes clear that reaching satellites with  $M_* \sim 10^4\, M_\odot$ should be enough to differentiate between models, since the ``power-law'' model predicts  almost $3$ times more such satellites than the ``cutoff'' model. The difference is most striking when reaching ultrafaints with $M_* \sim 10^2\; \rm M_\odot$, where only $10$ satellite dwarfs are expected around LMC-analogs in the case of a cutoff whereas more than $\sim 160$ are predicted for the ``power-law'' model. Should future surveys fail to discover a large number of ultrafaint dwarfs around isolated LMC-analogs, this would be strong evidence in favor of some kind of cutoff in the  stellar mass-halo mass relation.

\subsection{Comparison with previous work on satellites of LMC-like hosts}

It is interesting to compare our results with previous work in the literature on the satellite population of LMC-like hosts. For instance, assuming a power-law relation between stellar and halo mass, \citet{Nadler2020} predict $48 \pm 8$ satellites \footnote{\citet{Nadler2020} quote numbers above an absolute V-magnitude $M_V =0$, corresponding to a $M_* \sim 90 \rm M_\odot$ assuming a mass-to-light ratio $M/L=1$.} with $M_* \geq 10^2\; \rm M_\odot$, about a factor of three lower than the $\sim 160$ dwarf satellites predicted by our ``power-law'' model. This is not due to differences in our assumptions about the primary virial mass nor about the subhalo abundance: we have explicitly checked that the number of {\it subhalos} in our LMC-like primaries is consistent with \citet{Nadler2020}. Indeed, we find $48$-$58$ subhalos (10th-90th percentiles) with $V_{\rm peak} > 10$ km/s within the virial radius of primaries with $V_{200}=75$ km/s, in good agreement with the $52 \pm 8$ quoted by those authors. The difference must therefore be due to the way each model populates those subhalos with galaxies.

The slope in the low-mass end of the $M_*$-$V_{\rm peak}$ relation  
%assumed in
inferred by
 \citet{Nadler2020} is somewhat shallower than the one adopted here ($\sim 5.1$ compared to $\sim 7.4$ in our ``power-law'' model), but we have identified two main factors contributing to the smaller number of dwarf satellites predicted by \citet{Nadler2020} compared to our work. One is that  we model the scatter in the $M_*-V_{\rm peak}$ relation as velocity-dependent, increasing from $\sim 0.22$ dex for MW-like objects to $\sim 1$ dex in halos with $V_{\rm peak} \sim 10$ km/s. On the other hand, Fig.~6 (in combination with their Table~1) in \citet{Nadler2020} suggests that 
 %they assume 
their model infers 
 a roughly constant upper limit of $\sim 0.2$ dex scatter in the dwarf regime in order to reproduce the completeness-corrected number of observed MW satellites. 

The effect of the larger assumed scatter in our model is appreciable. Indeed, assuming zero scatter in the stellar mass - velocity relation, the ``power-law'' model would decrease the predicted numbers from $\sim 160$ to $\sim 67$ satellites with $M_* \geq 10^2\; \rm M_\odot$ (see middle dashed orange curve in Fig.~\ref{fig:modelfordelve}), in better agreement with the $48 \pm 8$ predicted in \citet{Nadler2020}. 
This is also in agreement with the $\sim 70$ satellites with $-7<M_V<-1$ predicted by \citet{Jethwa2016} via dynamical modelling of the Magellanic Cloud satellite population.
%\textcolor{red}{Moreover, assuming a scatter-free stellar mass - velocity relation that has the same slope as that in \citet{Nadler2020} results in $XX \pm XX$ satellites with $M_* \geq 10^2\; \rm M_\odot$, indicating that the scatter has a larger impact in the predictions than the slope of the abundance matching relation. }
%??I don't understand the line above??  
In summary, these results show qualitative consistency with \citet{Munshi2021} who find that a scatter that grows with halo mass or $V_{\rm peak}$ steepens the slope of the faint end of the resultant satellite mass function.
We refer the reader to \citet{GarrisonKimmel2017} for a detailed discussion of the degeneracies in the slope/scatter of abundance matching models and the expected number of dwarfs and to \citet{Munshi2019} for an example of how different sub-grid physics and resolution might impact the slope/scatter of the stellar - halo mass relation.

A second factor affecting the  number of ultrafaints in \citet{Nadler2020} is 
that their model infers
%the assumption of 
an occupation fraction
such
%Indeed, \citet{Nadler2020} assumes 
that below $V_{\rm peak} \sim 9$ km/s an increasing fraction of halos with decreasing $V_{\rm peak}$  remain dark and never host a galaxy (modeled according to their Eq. 3), while our ``power-law'' model assumes an occupation fraction equal to 1 at all $V_{\rm peak}$. We note that adding an occupation fraction to a power-law $M_*$-$V_{\rm peak}$ relation effectively makes it steeper and more comparable to the ``cutoff'' model, lowering the total number of predicted faint satellites.
%%ISS: Nadler gives in fig6(left) that galaxy occupation fraction decreases after Mpeak ~ 1e8Msun.  i pass this to vmax assuming mpeak=m200 NFW.

Our predictions may also be compared with the work of \citet{Dooley2017b}, who explored the satellite population of LMC-like hosts using (power-law) extrapolations of several abundance matching models, including that of \citet{Moster2013}. The main difference with our own ``power-law'' model is that they also include an occupation fraction to model the effects of reionization. As such, their predictions are more similar to our ``cutoff'' model, with $\sim 10$-$15$ (median, depending on which particular abundance matching relation) dwarf satellites with $M_* \geq 10^2\; \rm M_\odot$ within a $50$ kpc radius of their hosts.
%depending on which particular abundance matching relation. 
%??I don't understand the line above??  
These results are bracketed by the predictions of our ``cutoff'' model,  with $6$-$13$ (10th-90th percentiles), and our ``power-law'' model, with $145-178$ satellites, although our numbers are within a larger volume of $r_{\rm 200}\sim 107$ kpc (corresponding to $V_{200}=75$ km/s).
%Again, this analysis highlights the potential impact of observationally constraining the ultrafaint population of LMC-like dwarfs for our understanding of galaxy formation efficiency in low mass dark matter halos. 
%%?? ISS: our models for fixed v200=75 correspond to r200=107kpc.  we cannot know what is the number of sats in 50 kpc. ??

Our predictions in Fig.~\ref{fig:modelfordelve} might also inform other satellite searches around LMC-analogs in the field such as the LBT-SONG survey or MADCASH \citep{Carlin2021}. At least two faint satellites have been identified around the Magellanic dwarf NGC 628 \citep{Davis2021} surveying only a fraction of its inferred virial extension with the Large Binocular Telescope as well as the confirmation of DDO113 as (interacting) satellite of NGC 4214. Additionally, two dwarfs have been confirmed as satellites of the Magellanic analogs NGC 2403 and NGC 4214 with HST data for the Hyper Suprime Cam survey MADCASH. \citet{Muller2020} report, in addition, two candidate faint dwarfs possibly associated to NGC 24 in the Sculptor group using the Dark Energy Camera. As data continues to accumulate around dwarf primaries, the census of their satellite population is starting to emerge as the most promising and effective way to constrain the galaxy-halo connection at the low mass end.

\section{Summary and Conclusions}
\label{sec:conclu}

We have studied the effects of different stellar mass-halo mass relations on the predicted population of faint and ultrafaint dwarf satellites around primaries spanning a wide range of stellar mass. The models are motivated by results of recent state-of-the-art cosmological hydrodynamical simulations, extrapolated  to the ultrafaint regime, down to $M_* \sim 10^2\, M_\odot$. 

Two faint-end stellar mass-halo mass model relations are explored: one  is a ``power-law'' motivated by recent semianalytic results about the  abundance of satellites in the Local Group, and by recent high resolution simulations from the FIRE project, which follow closely a power-law extrapolation to the faint regime of the abundance-matching results from \citet{Moster2013}.

A second is a ``cutoff'' model where the stellar mass-halo mass relation gradually steepens towards decreasing mass so that no luminous dwarf exists beyond a minimum threshold halo virial mass of order $M_{200} \sim 10^{9}\, M_\odot$. The ``cutoff'' model is motivated by results from the APOSTLE simulations, and by analytic considerations that disfavour the formation of galaxies in halos below the ``hydrogen-cooling'' limit \citep[see; e.g.,][and references therein]{BenitezLlambay2020}. We assume the same subhalo mass function and the same mass-dependent scatter in the $M_*$-$V_{\rm peak}$ relation  in both models.
%All of these simulations have claimed successes at reproducing the observed properties of dwarf galaxies and MW-like satellite populations in the classical regime ($M_* \geq 10^5$ \msun).

Our main finding is that satellite mass functions of primary galaxies spanning a wide range of mass is an excellent probe of the shape of the stellar mass-halo mass relation at the faint end. Satellite mass functions are particularly constraining if they reach deep into the ultrafaint regime. For example, the ``cutoff'' model predicts $\sim 9(19)$ times fewer dwarfs with $M_* \geq 10^2$ \msun\; than the ``power-law'' model for primaries with $M_*\sim 3\times10^{10}( 3\times10^{7})$\msun. The difference becomes more marked as the stellar mass of the primary decreases, implying that the satellites of dwarf primaries, in particular, provide particularly strong constraints on the stellar mass-halo mass relation at the faintest end.

The models also predict different {\it normalized} satellite mass functions, i.e., the number of satellites expressed as a function of $M_*^{\rm sat}/M_*^{\rm pri}$ rather than $M_*^{\rm sat}$.  While the normalized function declines with decreasing primary stellar mass in the ``cutoff'' model, it is nearly independent of primary mass in the ``power-law'' model. This self-similar behavior results because the subhalo mass function is also a power-law in LCDM, as discussed by \citet[][]{Sales2013}.

Our findings have important implications when applied to nearby galaxies, where the surveying of ultrafaint dwarfs is or will become feasible in the near future. For a MW-mass primary (i.e., $V_{200} \sim 150$ km/s), the ``power-law'' and ``cutoff'' models predict $\sim 612^{+35}_{-25}$ vs. $\sim 77^{+10}_{-11}$ satellites above the nominal $M_* = 10^2\, M_\odot$ mass cut, respectively. Interestingly, in the MW itself the number of already discovered satellite dwarfs is quite close to the ``cutoff'' model prediction, leaving only little room for the detection of large numbers of new ultrafaint dwarfs. This is a prediction that should also be testable by ongoing and future surveys of the satellite population around nearby galaxies.

LCDM predicts that dwarf galaxies should also host a number of fainter companions. The models described above may be used to compute  the number of dwarf satellites expected around LMC-mass systems in the field. We find, on average, that the ``cutoff'' model predicts  $\sim 3$-$22$ satellites with $M_*>100\, M_\odot$; the number, on the other hand, grows to  $\sim 65$-$300$ for the power-law case, where the range corresponds to assuming a virial velocity range between $50$ and $100$ km/s for the LMC halo.
This highlights the potential for ultrafaint discoveries in regions surrounding Magellanic-like dwarfs in the field, a particularly exciting prospect in light of ongoing efforts such as DELVE \citep{Drlica2021}, MADCASH \citep{Carlin2021} or LBT-SONG \citep{Davis2021}, which target the surroundings of isolated LMC-like dwarfs. 

We conclude that the satellite population of dwarf galaxies in the field offer a powerful way to constrain the faint end of the stellar mass-halo mass relation \citep[see also][]{Wheeler2015}. Only if the relation extends well below the hydrogen-cooling limit (as envisioned in the ``power-law'' model) then one would expect dwarfs to have numerous ultrafaint companions. In the cutoff case, as the mass of the primary approaches the cutoff the number of satellites should decline rapidly. For the particular cutoff we explore in this paper, dwarfs with $M_* \sim 3\times 10^8\, M_\odot$ or less should have virtually no luminous satellites, regardless of luminosity. 

Dwarf primaries are also good probes because the galaxy mass is, in relative terms, much smaller that that of their surrounding halo. The galaxy's  effect on the subhalo population is therefore proportionally reduced compared to galaxies like the Milky Way, where the disk is massive enough to affect noticeably the evolution and survival of subhalos in its vicinity \citep{Jahn2019}.  Finally, dwarf primaries are more abundant in the Local Volume than giant  spirals like the MW or M31, so surveying a statistically meaningful sample becomes, in principle, more feasible.

We conclude that the satellite mass function of dwarf galaxies in the field represents an efficient and attractive approach for constraining the mapping between stars and dark matter halos in the low mass end with deliverables expected in the foreseeable future.

%While the sheer number of dwarf companions in primaries of a fixed stellar mass may help distinguish a power-law from a cutoff like model, the ultimate test will come from studying the satellite population of primaries in a wide range of masses, extending even below LMC-like hosts. As discussed above and highlighted in \citep{Sales2013}, identifying a lack of scaling with primary mass will be the clear fingerprint of a power-law behavior in the low mass end of the galaxy-halo relation. 

\section*{Acknowledgements}
ISS is supported by the Arthur B. McDonald Canadian Astroparticle Physics Research Institute. 
LVS is grateful for financial support from the NSF-CAREER-1945310, AST-2107993 and NASA- ATP-80NSSC20K0566 grants.
JFN is a Fellow of the Canadian Institute for Advanced Research. 
AF is supported by a UKRI Future Leaders Fellowship (grant no MR/T042362/1). 
This work used the DiRAC@Durham facility managed by the Institute for Computational Cosmology on behalf of the STFC DiRAC HPC Facility (www.dirac.ac.uk). The equipment was funded by BEIS capital funding via STFC capital grants ST/K00042X/1, ST/P002293/1, ST/R002371/1, and ST/S002502/1, Durham University and STFC operations' grant ST/R000832/1. DiRAC is part of the National e-Infrastructure.

%%%%%%%%%%%%%%%%%%%%%%%%%%%%%%%%%%%%%%%%%%%%%%%%%%
\section*{Data Availability}
The APOSTLE simulation data and model results underlying this article can be shared on reasonable request to the corresponding author. 
For data from other simulations we refer the interested reader to the corresponding references cited in each case.
The observational data for Local group satellites used in this article comes from 
 \citet[][see \url{http://www.astro.uvic.ca/~alan/Nearby_Dwarf_Database_files/NearbyGalaxies.dat}, and references therein]{McConnachie2012}.


\bibliographystyle{mnras}
\bibliography{archive} % if your bibtex file is called example.bib

\bsp	% typesetting comment
\label{lastpage}
\end{document}